\documentclass[twocolumn,showpacs,preprintnumbers,amsmath,amssymb,prl,superscriptaddress,longbibliography]{revtex4-1}
\usepackage{bbm}
\usepackage{mathrsfs}
\usepackage{graphicx}
\usepackage{dcolumn}
\usepackage{bm}
\usepackage{amsmath}
\usepackage{amsfonts}
\usepackage{color}
\usepackage[colorlinks=true,citecolor=blue,anchorcolor=cyan]{hyperref}
\usepackage{floatrow}

\begin{document}

\title{Dissipative Edge Transport in Disordered Axion Insulator Films}
\author{Zhaochen Liu}
\affiliation{State Key Laboratory of Surface Physics and Department of Physics, Fudan University, Shanghai 200433, China}
\author{Dongheng Qian}
\affiliation{State Key Laboratory of Surface Physics and Department of Physics, Fudan University, Shanghai 200433, China}
\author{Yadong Jiang}
\affiliation{State Key Laboratory of Surface Physics and Department of Physics, Fudan University, Shanghai 200433, China}
\author{Jing Wang}
\thanks{wjingphys@fudan.edu.cn}
\affiliation{State Key Laboratory of Surface Physics and Department of Physics, Fudan University, Shanghai 200433, China}
\affiliation{Institute for Nanoelectronic Devices and Quantum Computing, Fudan University, Shanghai 200433, China}
\affiliation{Zhangjiang Fudan International Innovation Center, Fudan University, Shanghai 201210, China}

\begin{abstract}
We investigate the role of disorder in the edge transport of axion insulator films. We predict by first-principles calculations that even-number-layer MnBi$_2$Te$_4$ have gapped helical edge states. The random potential will dramatically modify the edge spectral function to become \emph{gapless}. However, such gapless helical state here is fundamentally different from that in quantum spin Hall insulator or topological Anderson insulator. We further study the edge transport in this system by Landauer-B\"{u}ttiker formalism, and find such gapless edge state is dissipative and not immune to backscattering, which would explain the dissipative nonlocal transport in the axion insulator state observed in six septuple layer MnBi$_2$Te$_4$ experimentally. Several transport experiments are proposed to verify our theory on the dissipative helical edge channels. In particular, the longitudinal resistance can be greatly reduced by adding an extra floating probe even if it is not used. These results will facilitate the observsation of long-sought topological magnetoelectric effect in axion insulators.
\end{abstract}

\date{\today}


\maketitle

Topological phenomena have been one of the central topics in condensed matter physics~\cite{thouless1998,hasan2010,qi2011,tokura2019}. The interplay between band topology and magnetism gives rise to a variety of exotic quantum states~\cite{tokura2019,qi2008,wang2017c}. A prime example is the quantum anomalous Hall (QAH) effect discovered in magnetic topological insulator (TI) flims~\cite{chang2013b,checkelsky2014,kou2014,bestwick2015,chang2015,mogi2015,watanabe2019,deng2020}, where the spin-orbit coupling and ferromagnetic (FM) ordering combine to give rise to a topologically nontrivial phase characterized by a finite Chern number and gapless chiral edge states~\cite{qi2008,liu2008,yu2010,wang2013a}. Another interesting example is axion insulator, which is three-dimensional magnetic TI with a nonzero quantized Chern-Simons magnetoelectric coupling (axion $\theta=\pi$) protected by inversion symmetry $\mathcal{I}$ instead of time-reversal symmetry $\Theta$~\cite{qi2008,essin2009,coh2011,wan2012,turner2012,varnava2018,sekine2021}. Such axion coupling leads to the prediction of topological magnetoelectric (TME) effect~\cite{qi2008}, which is the hallmark of axion insulator but remains unexplored due to difficulties in realizing the axion insulator state.

The simplest scenario for axion insulator state is obtained in bulk TI with a surface gap induced by a hedgehog magnetization while preserving the bulk gap~\cite{qi2008,nomura2011,wang2015b,morimoto2015}. In the thin-film geometry, the above condition of hedgehog magnetization is simply fulfilled with an antiparallel magnetization on top and bottom surfaces, where the absence of all surface state transport leads to a zero Hall plateau $\sigma_{xy}=0$, $\sigma_{xx}\rightarrow0$, and $\rho_{xy}=0$, $\rho_{xy}\rightarrow\infty$~\cite{wang2014a,wang2015b}. Such peculiar charge transports have been observed in FM-TI-FM heterostructure~\cite{mogi2017,mogi2017a,xiao2018,grauer2017} and even layer MnBi$_2$Te$_4$ antiferromagnetic (AFM) TI~\cite{liu2020}, which were predicted to be axion insulator state~\cite{wang2015b,zhang2019,li2019,otrokov2019a}. Theoretically, the low-energy physics in two different systems are similar and generate topological $\theta$ response which is nonquantized due to finite-size effect~\cite{wang2015b,morimoto2015,liuzc2020a}. However, recent transport and microwave imaging experiments find quite different behaviors in these two systems, where gapless edge states do not exist in the former~\cite{mogi2017,allen2019} but do exist in the latter~\cite{li2021,lin2021}. Especially, the edge transport in MnBi$_2$Te$_4$ even layer is shown to be dissipative~\cite{li2021}. Thus, it is important to trace where such dissipative gapless edge states come from and understand the origin of the discrepancy in these two systems.

Here we study the role of disorder in the edge transport of axion insulator films. By combining first-principles calculations and analytic models, we show that six septuple layers (SL) of MnBi$_2$Te$_4$ studied in experiments~\cite{li2021,lin2021} have gapped helical edge states. A random potential will modify the edge spectral function to become gapless. Such gapless edge state is dissipative and not immune to backscattering, which would explain dissipative transport of the recent transport and image experiments~\cite{li2021,lin2021}.

\begin{figure}[t]
\begin{center}
\includegraphics[width=3.4in,clip=true]{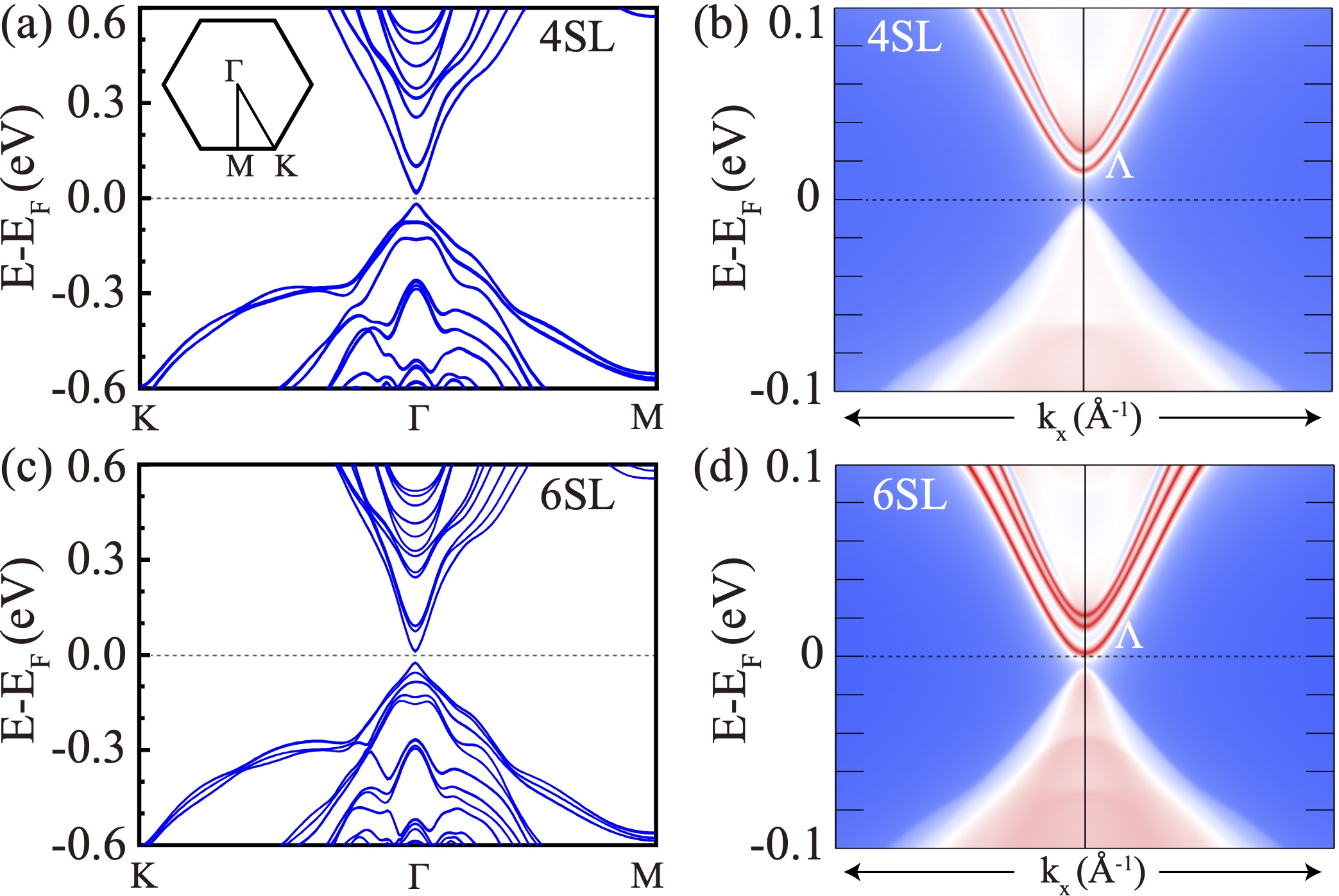}
\end{center}
\caption{(a) \& (c) Band structure for 4 SL and 6 SL MnBi$_2$Te$_4$. The dashed line indicates the Fermi level. The inset of (a) shows 2D Brillouin zone with high-symmetry \textbf{k} points $\Gamma$(0,0), $K$($\pi$,$\pi$) and $M$($\pi$,0) labelled. The energy dispersion of the semi-infinite film along edge $\Gamma M$ is plotted for (b) 4 SL and (d) 6 SL, respectively. The \emph{gapped} edge states are clearly seen around $\Gamma$ point as red lines dispersing in the 2D bulk gap.}
\label{fig1}
\end{figure}

\emph{Materials.-} We carry out first-principles calculations on MnBi$_2$Te$_4$ films. The material consists of Van der Waals coupled SL and develops $A$-type AFM order with an out-of-plane easy axis below N\'eel temperature, which is FM within each SL but AFM between adjacent SL along $z$ axis. The bulk state is an AFM TI with nontrivial $Z_2$ index protected by $S=\Theta\tau_{1/2}$~\cite{mong2010}, where $\tau_{1/2}$ is the half translation operator along $z$ axis. The odd SL with a net magnetization breaks $\mathcal{I}\Theta$ and shows QAH effect~\cite{zhang2019,li2019,otrokov2019a,deng2020}. While even SL with full compensated magnetic layers conserves $\mathcal{I}\Theta$ and exhibits zero-plateau QAH~\cite{liu2020}. We study the edge band structure of even SL along edge $\Gamma M$ direction. As shown in Fig.~\ref{fig1}, the 2D band structure has an inverted band gap at the $\Gamma$ point, and there indeed exists \emph{gapped} helical edge state $\Lambda$ in the insulating bulk. As we show below, it originates from helical edge states of the quantum spin Hall (QSH) effect but with $\Theta$-breaking due to magnetic ordering, where the gap is opened at Dirac point.

\emph{Model.-} The effective model for the low energy physics of even SL can be written down near the $\Gamma$ point. We start from 3D Hamiltonian $H_{\text{3d}}(\mathbf{k})$ for AFM MnBi$_2$Te$_4$, which is the same as that for $\Theta$-invariant TI due to conserved $\mathcal{S}$~\cite{zhang2019}. For even SL, $\mathcal{S}$ is broken and a term $H_{\text{ex}}$ describing the spatial alternating exchange field enters into $H_{\text{3d}}(\mathbf{k})$. The confinement in $z$ direction quantizes $k_z$ and leads to 2D subbands labeled by the subband index $n$. The 2D subbands have band inversion for film thickness $\geq$4 SL~\cite{zhang2019}, and without $H_{\text{ex}}$, the system is QSH with the low energy physics determined by Dirac surface states on top and bottom surfaces~\cite{liu2010a,lu2010}. The effect of $H_{\text{ex}}$ is to introduce opposite Zeeman terms on these two surfaces. Thus the effective model for even SL described by the massive Dirac surface states is given by~\cite{wang2015b,zhangjl2019,sun2020}
\begin{equation}\label{model}
\mathcal{H}_0(\mathbf{k})=\epsilon_0(\mathbf{k})+v(k_y\sigma_x-k_x\sigma_y)\tau_z+m(\mathbf{k})\tau_x+\Delta\sigma_z\tau_z,
\end{equation}
with the basis of $|t\uparrow\rangle,|t\downarrow\rangle,|b\uparrow\rangle$ and $|b\downarrow\rangle$, where $t$, $b$ denote top and bottom surfaces and $\uparrow$, $\downarrow$ represent spin up and down states, respectively. The particle-hole asymmetry $\epsilon_0(\mathbf{k})$ is neglected for simplicity. $\sigma_{i}$ and $\tau_{i}$ ($i=x,y,z$) are Pauli matrices acting on the  spin and layer, respectively. $v$ is the Dirac velocity, $m(\mathbf{k})=m_0+m_1(k_x^2+k_y^2)$ describes the tunneling effect between $t$ and $b$ surface states, $\Delta$ is the exchange field along $z$ axis introduced by the opposite magnetic ordering on $t$ and $b$. 

Eq.~(\ref{model}) correctly characterizes the gapped helical edge state shown in Fig.~\ref{fig1}. The energy gap for 2D bulk is $2\sqrt{m_0^2+\Delta^2}$ at $\Gamma$ point. If $\Delta=0$, this model is similar to Bernevig-Hughes-Zhang model for HgTe quantum wells~\cite{bernevig2006c} describing QSH with $m_0m_1<0$, where there exists gapless helical edge state. Then $\Delta$ further induces a gap to the edge state. The effective model for 1D gapped helical edge state is obtained analytically as $H_{\text{1d}}=vk_x\varrho_z+\Delta\varrho_x$, where $\varrho_i$ are Pauli matrices denoting pseudo-spin. The edge state gap $2\Delta$ is less than that of 2D bulk, consistent with Fig.~\ref{fig1}. It is worth mentioning there also exist other gapped helical edge states with higher energy than $\Lambda$ in 2D bulk  gap as shown in Fig.~\ref{fig1}(d), which are from the band inversion of extra 2D subbands with $n>1$ in thick film~\cite{supple}. In the following we investigate the edge transport determined by $\Lambda$ in the presence of disorder. Take 6 SL for a concrete example, we fit the parameters $v=3.2$~eV$\cdot\text{\AA}$,$m_0=-0.014$~eV, $m_1=9.4$~eV$\cdot\text{\AA}^2$ and $\Delta_z=5$~meV. 

In general, the disorder will generate spatially random perturbations to the pure Hamiltonian $\mathcal{H}_0$. Specifically, the system mainly has random scalar potential $\mathcal{H}_{U}=U(\mathbf{r})$ induced by impurities in the materials. There also exists random exchange field along $z$ axis induced by the inhomogenous AFM ordering $\mathcal{H}_{\Delta}=\Delta(\mathbf{r})\sigma_z\tau_z$. Here we are interested in system deep in AFM axion state and the fluctuation of $|\Delta(\mathbf{r})|<|\Delta|$, thus the random $\Delta(\mathbf{r})$ just renormalizes $\Delta$ to a reduced value in Eq.~(\ref{model}) and will not affect the edge transport essentially. Therefore we only need to consider $\mathcal{H}_U$, which is nonuniform and random in space but constant in time. 

\emph{Analysis of disorder.} 
Now we will show that disorder will renormalize Eq.~(\ref{model}). We extract the renormalized topological mass $m_0$, and the renormalized exchange field $\Delta$, from the self-energy $\Sigma$ of the disorder-averaged effective medium. In numerical simulations, we discretize $\mathcal{H}_0(\mathbf{k})$ on a square lattice and take a random on-site disorder potential $U(\mathbf{r})$, uniformly distributed in the interval $(-U_0,U_0)$. We denote $H_0(\mathbf{k})$ as the lattice Hamiltonian for Eq.~(\ref{model}).

\begin{figure}[t] 
\begin{center}
\includegraphics[width=3.4in,clip=true]{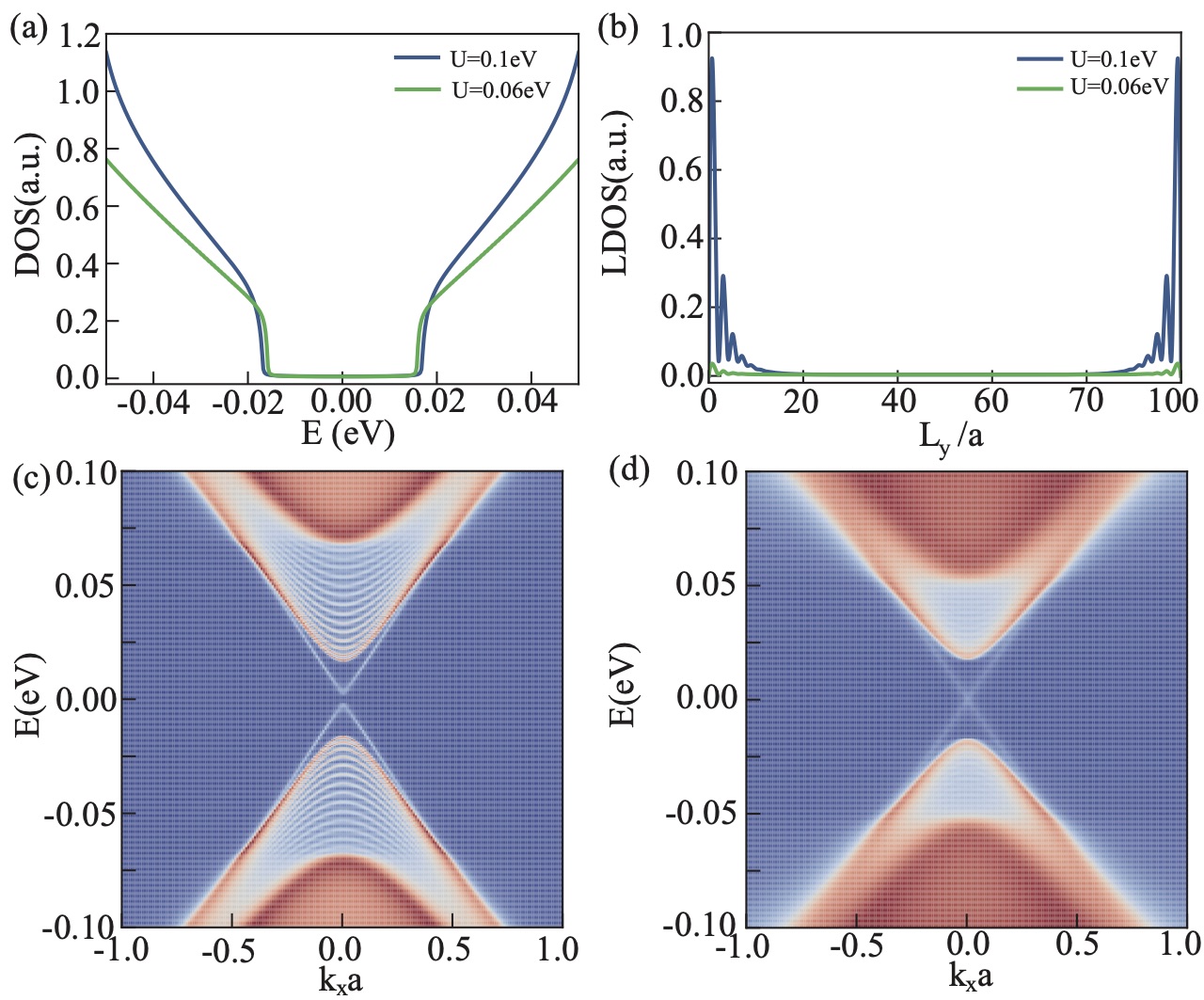}
\end{center} 
\caption{(a) DOS of 2D bulk for typical $U_0$. (b) The real space distribution for state at $E=0$ in (c) and (d). (c,d) The edge spectral function $\mathcal{A}(k,\omega)$ within SCBA of disorder strength $U_0=0.06$~eV and $U_0=0.1$~eV, respectively. A cylinder geometry is adopted with periodic boundary condition along $x$-axis and open boundary condition in $y$-axis with width $L_y=100a$. The lattice constant of the discretization $a=2$~nm.}
\label{fig2}
\end{figure}

The self-energy defined by $(E_F-H_0-\Sigma)^{-1}=\langle(E_F-H)^{-1}\rangle$, with $\langle...\rangle$ the disorder average, is a $4\times4$ matrix which we decompose into $\Gamma$ matrices: $\Sigma=\Sigma_0+\Sigma_1\sigma_x\tau_z+\Sigma_2\sigma_y\tau_z+\Sigma_4\tau_x+\Sigma_5\sigma_z\tau_z$. Then the renormalized $\widetilde{m}_0$ and $\widetilde{\Delta}$ are given by
\begin{equation}
\widetilde{m}_0=m_0+\text{Re}\Sigma_4,\ \ \widetilde{\Delta}=\Delta+\text{Re}\Sigma_5.
\end{equation}
The self-consistent Born approximation (SCBA) is employed to capture the main feature of disorder~\cite{altland2010}, where $\Sigma$ is given by the self-consistent equation,
\begin{equation}
\Sigma=\frac{U_0^2}{3}\left(\frac{a}{2\pi}\right)^2\int_{\text{BZ}}d^2\mathbf{k}\frac{1}{\omega-H_0(\mathbf{k})-\Sigma(\omega)+i0^+}.
\end{equation}
The self-energy is momentum independent, so there is no renormalization to $v$ and $m_1$. The corrections to $m_0$ and $\Delta$ are obtained approximately as
\begin{subequations}
\begin{align}
\widetilde{m}_0-m_0 &= -\frac{U^2_0a^2}{12\pi}\frac{1}{m_1}\ln\left|\frac{m_1^2\Pi^4}{m_0^2+\Delta^2-E_F^2}\right|,\label{m0}
\\
\widetilde{\Delta}-\Delta &= \frac{U^2_0a^2}{6\pi}\frac{\Delta\tanh^{-1}\left[\mathcal{Z}(k)\right]}{\sqrt{v^4+4v^2m_0m_1+4m_1^2(E_F^2-m_0^2)}}\Bigg|_{0}^{\Pi},\label{delta}
\\
\mathcal{Z}(k) &= \frac{v^2+2m_0m_1+2m_1^2k^2}{\sqrt{v^4+4v^2m_0m_1+4m_1^2(E_F^2-m_0^2)}},
\end{align}
\end{subequations}
where $\Pi=\pi/a$ is ultraviolet cutoff in momentum. Here we only keep the most logarithmically divergent term in Eq.~(\ref{m0}). The sign of $\widetilde{m}_0$ and $m_0$ remains the same, as the the correction to $m_0$ has opposite sign to $m_1$. Then the system is always in the inverted region~\cite{groth2009}. Similarly, the renormalized $\widetilde{\Delta}$ only decreases slightly. Therefore, the topological property of 2D bulk remains unchanged due to disorder, which is evidenced in the density-of-state (DOS) calculation in Fig.~\ref{fig2}(a).

To get information about the edge excitations in the disordered system, we further calculate the edge spectral function $\mathcal{A}(k,\omega)$ within SCBA in a cylinder geometry. The self-energy is $\Sigma(\omega,y)=(U_0^2/3)(a/2\pi)\int dk_x\mathcal{G}(\omega,k_x;y,y)$, with $\mathcal{G}(\omega,k_x;y,y)$ be the Green's function on cylinder, and the Dyson equation is $\mathcal{G}(k_x;y_1,y_0)=\int dy \mathcal{G}_0(k_x;y_1,y)\Sigma(y)\mathcal{G}(k_x;y,y_0)+\mathcal{G}_0(k_x;y_1,y_0)$. In a lattice $\int dy\rightarrow a\sum_i$, we have $\mathcal{G}(k_x)^{-1}=\mathcal{G}_0(k_x)^{-1}-\Sigma$. The spectral function $\mathcal{A}(k_x,\omega)=-(1/\pi)\text{Im}\mathcal{G}^R(k_x,\omega)$ is plotted in Fig.~\ref{fig2} for different disorder strength $U_0$. We can see that the disorder broadens the quasiparticle spectral weight and reduces the edge gap when $U_0$ is relatively small. While $U_0$ exceeds a critical value $U_c$, the edge spectrum is gapless as shown in Fig.~\ref{fig2}(d), and such gapless state indeeds resides at the sample boundary in Fig.~\ref{fig2}(b). This explains the gapless edge state observed in this system by microwave impedance microscopy~\cite{lin2021}. We point out that the gapless edge state here in the spectral function is essentially \emph{different} from that in topological Anderson insulator~(TAI)~\cite{li2009,groth2009,jiang2009}. In TAI, the gapless helical edge state is induced by disorder driven band inversion, which is dissipationless and immune from backscattering as protected by $\Theta$. Here in disordered axion insulator film, the edge state is dissipative because $\Theta$-breaking $\Delta$ induces backscattering. This is the main result of this paper.

The dissipative nature could be understood from effective theory for edge state with action 
\begin{equation}
\mathcal{S}=\int dtdx \psi^\dagger (\partial_t-iv\varrho_z\partial_x+\Delta\varrho_x+\mu(x))\psi,
\end{equation}
where $\mu(r)$ is the edge disorder potential with a zero mean. Via a nonlocal transformation $\psi=\mathcal{Q}(x)\widetilde{\psi}$ where $\mathcal{Q}(x)=\mathcal{P}\exp (-i\varrho_z\int^{x}_{-\infty}dx' \mu(x')/v)$, one can rewrite the action as
\begin{equation}\label{action}
\mathcal{S}=\int dtdx \widetilde{\psi}^\dagger (\partial_t-iv\varrho_z\partial_x+\Delta\mathcal{Q}^\dag\varrho_x\mathcal{Q})\psi
\end{equation}
where $\mathcal{P}$ stands for path ordering. The last term in Eq.~(\ref{action}) has long range correlation from the random string phase factor $\mathcal{Q}$, which is a relevant perturbation and describes backscattering. This term is absent in the quantum Hall chiral edge states with $\nu=2$ filling due to $SU(2)$ symmetry~\cite{kane2007,giamarchi2003}. Now we can see that the  transformed action describes a gapless helical edge state with a backscattering term from random disorder. 

\begin{figure}[t] 
\begin{center}
\includegraphics[width=3.4in,clip=true]{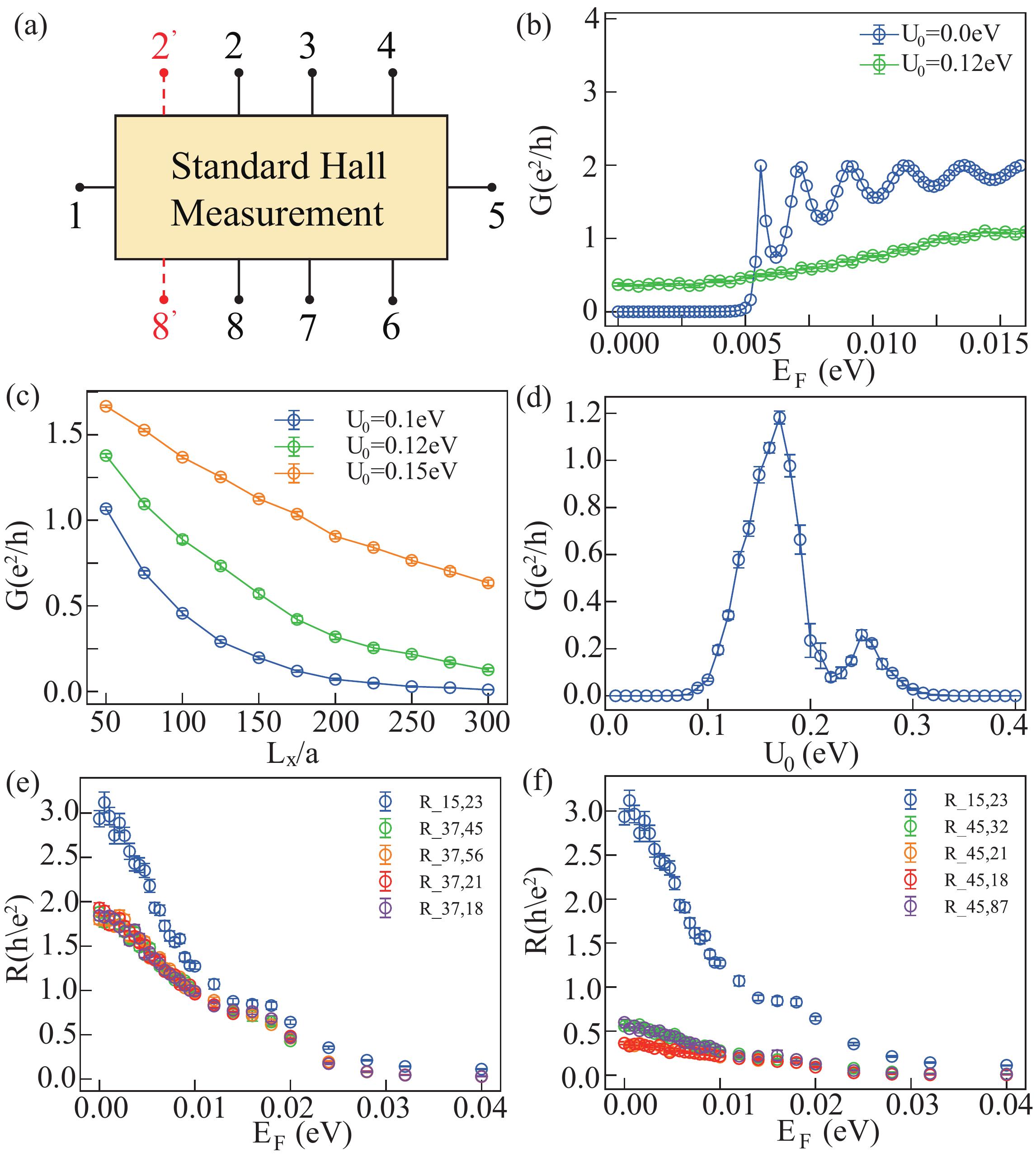}
\end{center} 
\caption{(a) Schematic drawing of a Hall bar device. (b) The two terminal conductance $G$ vs $E_F$ with different $U_0$. (c) $G$ vs $L_x$ at $E_F=0$ for different $U_0$, where $L_y=100a$. (d) $G$ vs $U_0$ at $E_F=0$. The device size in (b), (d) is $L_x\times L_y=200a\times100a$. (e,f) The local and nonlocal resistance $R_{ij,kl}$ in eight terminal device as a function of $E_F$ with $U_0=0.1$~eV. The device size is $L_x\times L_y=300a\times200a$. Each data point is a result of averaging over 500 disorder realizations.}
\label{fig3}
\end{figure} 

\emph{Numerics.} 
The above analytic results can be corroborated numerically by using the package Kwant~\cite{groth2014}. The resistance is calculated by the Landauer-B\"uttiker formalism with disorder-averaged transmission amplitude. The device geometry with standard Hall bar is illustrated in Fig.~\ref{fig3}(a). The two terminal conductance $G$ as a function of Fermi energy $E_F$ is shown in Fig.~\ref{fig3}(b). In the clean limit, $G$ vanishes when $E_F$ is in the edge gap and is finite exhibiting oscillating behaviour when $E_F$ is in the $\Lambda$ band, where the transmission resonance $G=2e^2/h$ is consistent with the gapped helical edge state. For finite disorder, $G$ is finite when $E_F$ is even in the edge gap (of the clean limit) and gradually grows as $E_F$ increases. The disappearance of conductance oscillation and $G<2e^2/h$ when $E_F$ is in the conducting edge band are the manifestation of dissipative nature of edge state. $G$ as a function of disorder strength $U_0$ at $E_F=0$ is plotted in Fig.~\ref{fig3}(d). We can see $G$ is finite only with moderate $U_0$. When $U_0<U_c$, $G=0$ due to finite edge gap in the spectral function in Fig.~\ref{fig2}, while $G$ vanishes for strong $U_0$ is from the Anderson localization. Furthermore, the dissipative edge transport leads to monotonically decreasing $G$ versus increasing device length $L_x$ in Fig.~\ref{fig3}(c).

The dissipative transport measured in the two terminals does not allow us to distinguish experimentally between helical edge channels and residual bulk conduction channels in a convincing manner. An unambiguous way to reveal the existence of dissipative helical edge state transport in the system is to use nonlocal electrical measurements. The edge states necessarily lead to nonlocal transport, and such nonlocal transport provides definitive evidence for the existence of chiral edge states in the quantum Hall effect~\cite{buttiker1986,buttiker1988}. The nonlocal resistance $R_{ij,kl}$ is plotted in Fig.~\ref{fig3}(e) and~\ref{fig3}(f), which is defined as voltage between electrode $k$ and $l$ divided by the current flowing through electrode $i$ and $j$, i.e., $R_{ij,kl}=V_{kl}/I_{ij}$. All of the nonlocal resistances are greater than the corresponding quantized value for dissipationless gapless helical edge state in QSH, which further demonstrates the edge transport is dissipative here. The nonlocal resistances decreases and finally vanishes when $E_F$ further goes into bulk. Moreover, one interesting feature in Fig.~\ref{fig3}(f) is that $R_{15,23}\approx 4R_{45,kl}$, which agrees with the recent transport  experiment qualitatively~\cite{li2021}. We emphasize the transmission amplitude and resistance in numerical simulation depend on system size and position of electrodes, which is the key feature for dissipative edge transport in this system.

\emph{Edge transport.}
We further propose a theory for the dissipative edge transport within the general Landauer-B\"{u}ttiker formalism~\cite{buttiker1986,buttiker1988}, where the current-voltage relationship is expressed as $I_i=(e^2/h)\sum_{j}\left(T_{ji}V_i-T_{ij}V_j\right)$, where $V_i$ is the voltage on the $i$th electrode, $I_i$ is the current flowing out of the $i$th electrode into the sample, and $T_{ji}$ is the transmission probability from the $i$th to the $j$th electrode. There is no net current ($I_j=0$) on a voltage lead or floating probe $j$, and the total current is conserved, namely $\sum_iI_i=0$. The current is zero when all the potentials are equal, implying the sum rules $\sum_iT_{ji}=\sum_iT_{ij}$.

For a standard Hall bar with $\mathcal{N}$ current and voltage leads [such as Fig.~\ref{fig3}(a) with $\mathcal{N}=8$], the transmission matrix elements for the dissipative helical state are given by $T_{i+1,i}=T_{i,i+1}=\kappa_i$ (from the disorder-averaged $\mathcal{I}\Theta$ symmetry) and others $=0$ (Here we identify $i=\mathcal{N}+1$ with $i=1$). These states are not protected from backscattering and the transmission from one electrode to the next is not perfect, implying $\kappa_i<1$~\cite{wang2013b}, which is different from dissipationless helical edge states in QSH where $\kappa_i=1$~\cite{roth2009}. In general, $\kappa_i$ become zero for infinitely large sample, because dissipation occurs once the phase coherence is destroyed in the metallic leads or the momentum is relaxed $\kappa_i\sim e^{-\ell/l_\mathrm{m}}$, where $\ell$ is the size between adjacent leads, $l_{\mathrm{m}}$ is the mean free path which is $1/2$ of the localization length for 1D state~\cite{thouless1973}. For simplicity, we have assumed $T_{ij}$ to be translational invariant, namely $T_{i+1,i}=\kappa$ is $i$ independent. The edge theory leads to the two terminal conductance $G\sim \kappa e^2/h\propto e^{-L_x/l_\mathrm{m}}$, which agrees with Fig.~\ref{fig3}(c) quantitatively. Considering again the nonlocal transport as in Fig.~\ref{fig3}(f), one finds that $R_{15,23}=h/2\kappa e^2$, $R_{45,kl}=h/8\kappa e^2$, and the relation $R_{15,23}=4R_{45,kl}$.

\begin{figure}[t] 
\begin{center}
\includegraphics[width=3.4in,clip=true]{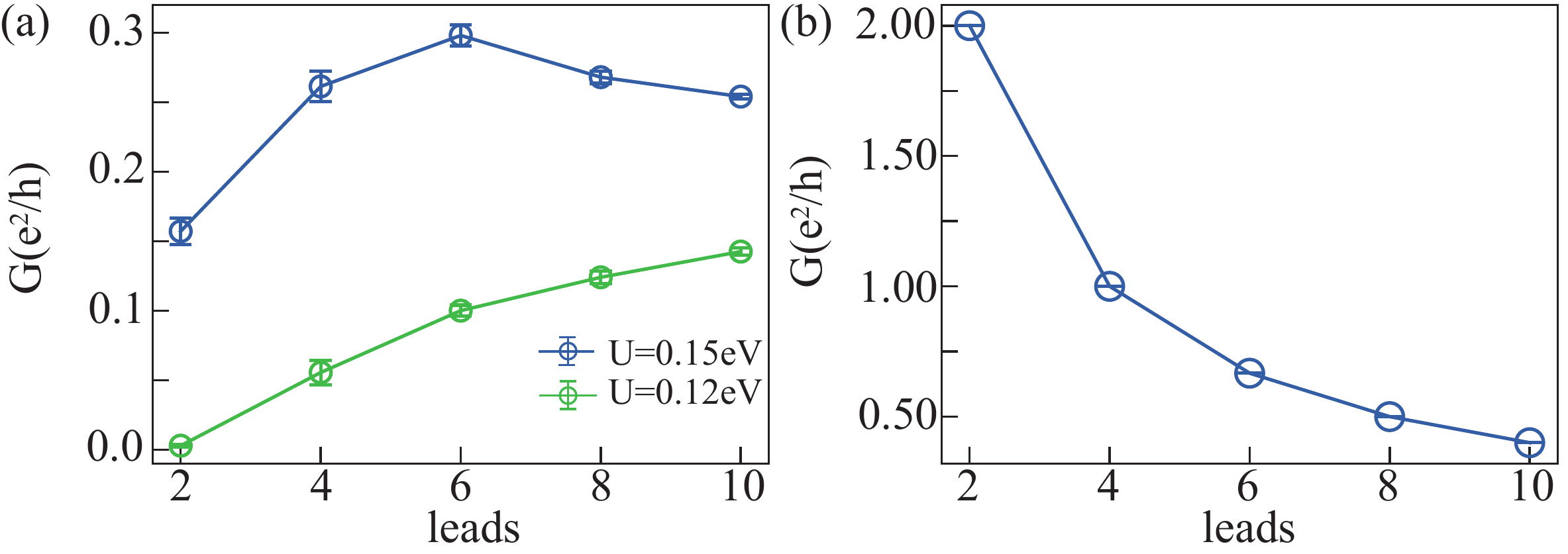}
\end{center} 
\caption{The longitudinal conductance vs number of floating leads for (a) dissipative helical state in 6 SL MnBi$_2$Te$_4$ with $U_0=0.12$~eV and $E_F=0$, and (b) dissipationless helical state in QSH. $L_x\times L_y=600a\times100a$.}
\label{fig4}
\end{figure} 

The effect of decoherence between two real leads can be modeled as an extra floating lead, in which dissipative gapless helical states interact with infinitely many low-energy degrees of freedom, completely losing their phase coherence~\cite{roth2009}. $\kappa$ is length dependent for the dissipative helical state in axion insulator film, while it is length independent ($\kappa=1$) for dissipationless helical state in QSH. This leads to quite different transport signatures between these two helical states. For example, if we put extra pairs of floating probes ($2'$ and $8'$ in Fig.~\ref{fig3}(a)) in the standard two terminal device with $L\gg l_{\mathrm{m}}$, we can see the longitudinal conductance increases (but not necessarily monotonically) as the number of floating leads increases for dissipative helical state in Fig.~\ref{fig4}(a)~\cite{supple}, which is just the opposite for QSH in Fig.~\ref{fig4}(b). This is a rather sharp feature which is easy to implement in experiments.

\emph{Discussion.-} 
The dissipative gapless helical edge state from disorder in MnBi$_2$Te$_4$ films and its transport properties well explain the recent transport and image experiments~\cite{li2021,lin2021}. The nonlocal resistance $R_{37,21}$ is greater than $R_{37,45}$,$R_{37,56}$ and $R_{37,18}$ in experiment~\cite{li2021}, since $\kappa$ is length dependent, one possible explanation is that the position of the electrodes are neither equally spaced nor perfectly aligned, which is common in experiments. Moreover, Eq.~(\ref{model}) also describes the low energy physics in FM-TI-FM heterostructure with $m_0\approx0$~\cite{mogi2017}, the disorder will induce band inversion with a negative renormalized $\widetilde{m}_0$. However, the disorder strength is expected to be small and the exchange field is large in such a modulated doping system~\cite{mogi2015}, thus the system should not have any gapless edge states. Finally, high magnetic field drives MnBi$_2$Te$_4$ even layer into Chern insulator state with a full magnetization. The helical edge state evolves into gapless chiral edge state, while other higher-energy helical states become quasi-helical states with a larger gap due to stronger exchange field, and the transport is only determined by the dissipationless chiral edge channel.

In summary, disorder with moderate strength will dramatically modify the edge transport in axion insulator films, which is a generic phenomenon. Thinner films of axion insulator such as 4 SL MnBi$_2$Te$_4$ has a larger edge gap as shown in Fig.~\ref{fig1}(b), such gapped state may persist even in the presence of disorder, and one can realize the long-sought TME effect in axion insulator without any gapless states.

\begin{acknowledgments}
We thank Jinsong Zhang and Eric Yue Ma for valuable discussions. This work is supported by the National Key Research Program of China under Grant No.~2019YFA0308404, the Natural Science Foundation of China through Grant Nos.~11774065 and~12174066, Shanghai Municipal Science and Technology Major Project under Grant No.~2019SHZDZX01, Science and Technology Commission of Shanghai Municipality under Grant No.~20JC1415900, and the Natural Science Foundation of Shanghai under Grant No.~19ZR1471400. Z.L. and D.Q. contributed equally to this work.
\end{acknowledgments}


\begin{thebibliography}{58}%
    \makeatletter
    \providecommand \@ifxundefined [1]{%
     \@ifx{#1\undefined}
    }%
    \providecommand \@ifnum [1]{%
     \ifnum #1\expandafter \@firstoftwo
     \else \expandafter \@secondoftwo
     \fi
    }%
    \providecommand \@ifx [1]{%
     \ifx #1\expandafter \@firstoftwo
     \else \expandafter \@secondoftwo
     \fi
    }%
    \providecommand \natexlab [1]{#1}%
    \providecommand \enquote  [1]{``#1''}%
    \providecommand \bibnamefont  [1]{#1}%
    \providecommand \bibfnamefont [1]{#1}%
    \providecommand \citenamefont [1]{#1}%
    \providecommand \href@noop [0]{\@secondoftwo}%
    \providecommand \href [0]{\begingroup \@sanitize@url \@href}%
    \providecommand \@href[1]{\@@startlink{#1}\@@href}%
    \providecommand \@@href[1]{\endgroup#1\@@endlink}%
    \providecommand \@sanitize@url [0]{\catcode `\\12\catcode `\$12\catcode
      `\&12\catcode `\#12\catcode `\^12\catcode `\_12\catcode `\%12\relax}%
    \providecommand \@@startlink[1]{}%
    \providecommand \@@endlink[0]{}%
    \providecommand \url  [0]{\begingroup\@sanitize@url \@url }%
    \providecommand \@url [1]{\endgroup\@href {#1}{\urlprefix }}%
    \providecommand \urlprefix  [0]{URL }%
    \providecommand \Eprint [0]{\href }%
    \providecommand \doibase [0]{http://dx.doi.org/}%
    \providecommand \selectlanguage [0]{\@gobble}%
    \providecommand \bibinfo  [0]{\@secondoftwo}%
    \providecommand \bibfield  [0]{\@secondoftwo}%
    \providecommand \translation [1]{[#1]}%
    \providecommand \BibitemOpen [0]{}%
    \providecommand \bibitemStop [0]{}%
    \providecommand \bibitemNoStop [0]{.\EOS\space}%
    \providecommand \EOS [0]{\spacefactor3000\relax}%
    \providecommand \BibitemShut  [1]{\csname bibitem#1\endcsname}%
    \let\auto@bib@innerbib\@empty
    \bibitem [{\citenamefont {Thouless}(1998)}]{thouless1998}%
      \BibitemOpen
      \bibfield  {author} {\bibinfo {author} {\bibfnamefont {D.~J.}\ \bibnamefont
      {Thouless}},\ }\href@noop {} {\emph {\bibinfo {title} {Topological Quantum
      Numbers in Nonrealistic Physics}}}\ (\bibinfo  {publisher} {World Scientific,
      Singapore},\ \bibinfo {year} {1998})\BibitemShut {NoStop}%
    \bibitem [{\citenamefont {Hasan}\ and\ \citenamefont {Kane}(2010)}]{hasan2010}%
      \BibitemOpen
      \bibfield  {author} {\bibinfo {author} {\bibfnamefont {M.~Z.}\ \bibnamefont
      {Hasan}}\ and\ \bibinfo {author} {\bibfnamefont {C.~L.}\ \bibnamefont
      {Kane}},\ }\bibfield  {title} {\enquote {\bibinfo {title}
      {\textit{Colloquium}: Topological insulators},}\ }\href {\doibase
      10.1103/RevModPhys.82.3045} {\bibfield  {journal} {\bibinfo  {journal} {Rev.
      Mod. Phys.}\ }\textbf {\bibinfo {volume} {82}},\ \bibinfo {pages}
      {3045--3067} (\bibinfo {year} {2010})}\BibitemShut {NoStop}%
    \bibitem [{\citenamefont {Qi}\ and\ \citenamefont {Zhang}(2011)}]{qi2011}%
      \BibitemOpen
      \bibfield  {author} {\bibinfo {author} {\bibfnamefont {Xiao-Liang}\
      \bibnamefont {Qi}}\ and\ \bibinfo {author} {\bibfnamefont {Shou-Cheng}\
      \bibnamefont {Zhang}},\ }\bibfield  {title} {\enquote {\bibinfo {title}
      {Topological insulators and superconductors},}\ }\href {\doibase
      10.1103/RevModPhys.83.1057} {\bibfield  {journal} {\bibinfo  {journal} {Rev.
      Mod. Phys.}\ }\textbf {\bibinfo {volume} {83}},\ \bibinfo {pages}
      {1057--1110} (\bibinfo {year} {2011})}\BibitemShut {NoStop}%
    \bibitem [{\citenamefont {Tokura}\ \emph {et~al.}(2019)\citenamefont {Tokura},
      \citenamefont {Yasuda},\ and\ \citenamefont {Tsukazaki}}]{tokura2019}%
      \BibitemOpen
      \bibfield  {author} {\bibinfo {author} {\bibfnamefont {Yoshinori}\
      \bibnamefont {Tokura}}, \bibinfo {author} {\bibfnamefont {Kenji}\
      \bibnamefont {Yasuda}}, \ and\ \bibinfo {author} {\bibfnamefont {Atsushi}\
      \bibnamefont {Tsukazaki}},\ }\bibfield  {title} {\enquote {\bibinfo {title}
      {Magnetic topological insulators},}\ }\href {\doibase
      10.1038/s42254-018-0011-5} {\bibfield  {journal} {\bibinfo  {journal} {Nat.
      Rev. Phys.}\ }\textbf {\bibinfo {volume} {1}},\ \bibinfo {pages} {126--143}
      (\bibinfo {year} {2019})}\BibitemShut {NoStop}%
    \bibitem [{\citenamefont {Qi}\ \emph {et~al.}(2008)\citenamefont {Qi},
      \citenamefont {Hughes},\ and\ \citenamefont {Zhang}}]{qi2008}%
      \BibitemOpen
      \bibfield  {author} {\bibinfo {author} {\bibfnamefont {Xiao-Liang}\
      \bibnamefont {Qi}}, \bibinfo {author} {\bibfnamefont {Taylor~L.}\
      \bibnamefont {Hughes}}, \ and\ \bibinfo {author} {\bibfnamefont {Shou-Cheng}\
      \bibnamefont {Zhang}},\ }\bibfield  {title} {\enquote {\bibinfo {title}
      {Topological field theory of time-reversal invariant insulators},}\ }\href
      {\doibase 10.1103/PhysRevB.78.195424} {\bibfield  {journal} {\bibinfo
      {journal} {Phys. Rev. B}\ }\textbf {\bibinfo {volume} {78}},\ \bibinfo
      {pages} {195424} (\bibinfo {year} {2008})}\BibitemShut {NoStop}%
    \bibitem [{\citenamefont {Wang}\ and\ \citenamefont {Zhang}(2017)}]{wang2017c}%
      \BibitemOpen
      \bibfield  {author} {\bibinfo {author} {\bibfnamefont {Jing}\ \bibnamefont
      {Wang}}\ and\ \bibinfo {author} {\bibfnamefont {Shou-Cheng}\ \bibnamefont
      {Zhang}},\ }\bibfield  {title} {\enquote {\bibinfo {title} {Topological
      states of condensed matter},}\ }\href {\doibase 10.1038/NMAT5012} {\bibfield
      {journal} {\bibinfo  {journal} {Nature Mat.}\ }\textbf {\bibinfo {volume}
      {16}},\ \bibinfo {pages} {1062--1067} (\bibinfo {year} {2017})}\BibitemShut
      {NoStop}%
    \bibitem [{\citenamefont {Chang}\ \emph {et~al.}(2013)\citenamefont {Chang},
      \citenamefont {Zhang}, \citenamefont {Feng}, \citenamefont {Shen},
      \citenamefont {Zhang}, \citenamefont {Guo}, \citenamefont {Li}, \citenamefont
      {Ou}, \citenamefont {Wei}, \citenamefont {Wang}, \citenamefont {Ji},
      \citenamefont {Feng}, \citenamefont {Ji}, \citenamefont {Chen}, \citenamefont
      {Jia}, \citenamefont {Dai}, \citenamefont {Fang}, \citenamefont {Zhang},
      \citenamefont {He}, \citenamefont {Wang}, \citenamefont {Lu}, \citenamefont
      {Ma},\ and\ \citenamefont {Xue}}]{chang2013b}%
      \BibitemOpen
      \bibfield  {author} {\bibinfo {author} {\bibfnamefont {Cui-Zu}\ \bibnamefont
      {Chang}}, \bibinfo {author} {\bibfnamefont {Jinsong}\ \bibnamefont {Zhang}},
      \bibinfo {author} {\bibfnamefont {Xiao}\ \bibnamefont {Feng}}, \bibinfo
      {author} {\bibfnamefont {Jie}\ \bibnamefont {Shen}}, \bibinfo {author}
      {\bibfnamefont {Zuocheng}\ \bibnamefont {Zhang}}, \bibinfo {author}
      {\bibfnamefont {Minghua}\ \bibnamefont {Guo}}, \bibinfo {author}
      {\bibfnamefont {Kang}\ \bibnamefont {Li}}, \bibinfo {author} {\bibfnamefont
      {Yunbo}\ \bibnamefont {Ou}}, \bibinfo {author} {\bibfnamefont {Pang}\
      \bibnamefont {Wei}}, \bibinfo {author} {\bibfnamefont {Li-Li}\ \bibnamefont
      {Wang}}, \bibinfo {author} {\bibfnamefont {Zhong-Qing}\ \bibnamefont {Ji}},
      \bibinfo {author} {\bibfnamefont {Yang}\ \bibnamefont {Feng}}, \bibinfo
      {author} {\bibfnamefont {Shuaihua}\ \bibnamefont {Ji}}, \bibinfo {author}
      {\bibfnamefont {Xi}~\bibnamefont {Chen}}, \bibinfo {author} {\bibfnamefont
      {Jinfeng}\ \bibnamefont {Jia}}, \bibinfo {author} {\bibfnamefont
      {Xi}~\bibnamefont {Dai}}, \bibinfo {author} {\bibfnamefont {Zhong}\
      \bibnamefont {Fang}}, \bibinfo {author} {\bibfnamefont {Shou-Cheng}\
      \bibnamefont {Zhang}}, \bibinfo {author} {\bibfnamefont {Ke}~\bibnamefont
      {He}}, \bibinfo {author} {\bibfnamefont {Yayu}\ \bibnamefont {Wang}},
      \bibinfo {author} {\bibfnamefont {Li}~\bibnamefont {Lu}}, \bibinfo {author}
      {\bibfnamefont {Xu-Cun}\ \bibnamefont {Ma}}, \ and\ \bibinfo {author}
      {\bibfnamefont {Qi-Kun}\ \bibnamefont {Xue}},\ }\bibfield  {title} {\enquote
      {\bibinfo {title} {{Experimental Observation of the Quantum Anomalous Hall
      Effect in a Magnetic Topological Insulator}},}\ }\href {\doibase
      10.1126/science.1234414} {\bibfield  {journal} {\bibinfo  {journal}
      {Science}\ }\textbf {\bibinfo {volume} {340}},\ \bibinfo {pages} {167--170}
      (\bibinfo {year} {2013})}\BibitemShut {NoStop}%
    \bibitem [{\citenamefont {Checkelsky}\ \emph {et~al.}(2014)\citenamefont
      {Checkelsky}, \citenamefont {Yoshimi}, \citenamefont {Tsukazaki},
      \citenamefont {Takahashi}, \citenamefont {Kozuka}, \citenamefont {Falson},
      \citenamefont {Kawasaki},\ and\ \citenamefont {Tokura}}]{checkelsky2014}%
      \BibitemOpen
      \bibfield  {author} {\bibinfo {author} {\bibfnamefont {J.~G.}\ \bibnamefont
      {Checkelsky}}, \bibinfo {author} {\bibfnamefont {R.}~\bibnamefont {Yoshimi}},
      \bibinfo {author} {\bibfnamefont {A.}~\bibnamefont {Tsukazaki}}, \bibinfo
      {author} {\bibfnamefont {K.~S.}\ \bibnamefont {Takahashi}}, \bibinfo {author}
      {\bibfnamefont {Y.}~\bibnamefont {Kozuka}}, \bibinfo {author} {\bibfnamefont
      {J.}~\bibnamefont {Falson}}, \bibinfo {author} {\bibfnamefont
      {M.}~\bibnamefont {Kawasaki}}, \ and\ \bibinfo {author} {\bibfnamefont
      {Y.}~\bibnamefont {Tokura}},\ }\bibfield  {title} {\enquote {\bibinfo {title}
      {Trajectory of the anomalous hall effect towards the quantized state in a
      ferromagnetic topological insulator},}\ }\href {\doibase 10.1038/nphys3053}
      {\bibfield  {journal} {\bibinfo  {journal} {Nature Phys.}\ }\textbf {\bibinfo
      {volume} {10}},\ \bibinfo {pages} {731} (\bibinfo {year} {2014})}\BibitemShut
      {NoStop}%
    \bibitem [{\citenamefont {Kou}\ \emph {et~al.}(2014)\citenamefont {Kou},
      \citenamefont {Guo}, \citenamefont {Fan}, \citenamefont {Pan}, \citenamefont
      {Lang}, \citenamefont {Jiang}, \citenamefont {Shao}, \citenamefont {Nie},
      \citenamefont {Murata}, \citenamefont {Tang}, \citenamefont {Wang},
      \citenamefont {He}, \citenamefont {Lee}, \citenamefont {Lee},\ and\
      \citenamefont {Wang}}]{kou2014}%
      \BibitemOpen
      \bibfield  {author} {\bibinfo {author} {\bibfnamefont {Xufeng}\ \bibnamefont
      {Kou}}, \bibinfo {author} {\bibfnamefont {Shih-Ting}\ \bibnamefont {Guo}},
      \bibinfo {author} {\bibfnamefont {Yabin}\ \bibnamefont {Fan}}, \bibinfo
      {author} {\bibfnamefont {Lei}\ \bibnamefont {Pan}}, \bibinfo {author}
      {\bibfnamefont {Murong}\ \bibnamefont {Lang}}, \bibinfo {author}
      {\bibfnamefont {Ying}\ \bibnamefont {Jiang}}, \bibinfo {author}
      {\bibfnamefont {Qiming}\ \bibnamefont {Shao}}, \bibinfo {author}
      {\bibfnamefont {Tianxiao}\ \bibnamefont {Nie}}, \bibinfo {author}
      {\bibfnamefont {Koichi}\ \bibnamefont {Murata}}, \bibinfo {author}
      {\bibfnamefont {Jianshi}\ \bibnamefont {Tang}}, \bibinfo {author}
      {\bibfnamefont {Yong}\ \bibnamefont {Wang}}, \bibinfo {author} {\bibfnamefont
      {Liang}\ \bibnamefont {He}}, \bibinfo {author} {\bibfnamefont {Ting-Kuo}\
      \bibnamefont {Lee}}, \bibinfo {author} {\bibfnamefont {Wei-Li}\ \bibnamefont
      {Lee}}, \ and\ \bibinfo {author} {\bibfnamefont {Kang~L.}\ \bibnamefont
      {Wang}},\ }\bibfield  {title} {\enquote {\bibinfo {title} {Scale-invariant
      quantum anomalous hall effect in magnetic topological insulators beyond the
      two-dimensional limit},}\ }\href {\doibase 10.1103/PhysRevLett.113.137201}
      {\bibfield  {journal} {\bibinfo  {journal} {Phys. Rev. Lett.}\ }\textbf
      {\bibinfo {volume} {113}},\ \bibinfo {pages} {137201} (\bibinfo {year}
      {2014})}\BibitemShut {NoStop}%
    \bibitem [{\citenamefont {Bestwick}\ \emph {et~al.}(2015)\citenamefont
      {Bestwick}, \citenamefont {Fox}, \citenamefont {Kou}, \citenamefont {Pan},
      \citenamefont {Wang},\ and\ \citenamefont {Goldhaber-Gordon}}]{bestwick2015}%
      \BibitemOpen
      \bibfield  {author} {\bibinfo {author} {\bibfnamefont {A.~J.}\ \bibnamefont
      {Bestwick}}, \bibinfo {author} {\bibfnamefont {E.~J.}\ \bibnamefont {Fox}},
      \bibinfo {author} {\bibfnamefont {Xufeng}\ \bibnamefont {Kou}}, \bibinfo
      {author} {\bibfnamefont {Lei}\ \bibnamefont {Pan}}, \bibinfo {author}
      {\bibfnamefont {Kang~L.}\ \bibnamefont {Wang}}, \ and\ \bibinfo {author}
      {\bibfnamefont {D.}~\bibnamefont {Goldhaber-Gordon}},\ }\bibfield  {title}
      {\enquote {\bibinfo {title} {Precise quantization of the anomalous hall
      effect near zero magnetic field},}\ }\href {\doibase
      10.1103/PhysRevLett.114.187201} {\bibfield  {journal} {\bibinfo  {journal}
      {Phys. Rev. Lett.}\ }\textbf {\bibinfo {volume} {114}},\ \bibinfo {pages}
      {187201} (\bibinfo {year} {2015})}\BibitemShut {NoStop}%
    \bibitem [{\citenamefont {Chang}\ \emph {et~al.}(2015)\citenamefont {Chang},
      \citenamefont {Zhao}, \citenamefont {Kim}, \citenamefont {Zhang},
      \citenamefont {Assaf}, \citenamefont {Heiman}, \citenamefont {Zhang},
      \citenamefont {Liu}, \citenamefont {Chan},\ and\ \citenamefont
      {Moodera}}]{chang2015}%
      \BibitemOpen
      \bibfield  {author} {\bibinfo {author} {\bibfnamefont {Cui-Zu}\ \bibnamefont
      {Chang}}, \bibinfo {author} {\bibfnamefont {Weiwei}\ \bibnamefont {Zhao}},
      \bibinfo {author} {\bibfnamefont {Duk~Y.}\ \bibnamefont {Kim}}, \bibinfo
      {author} {\bibfnamefont {Haijun}\ \bibnamefont {Zhang}}, \bibinfo {author}
      {\bibfnamefont {Badih~A.}\ \bibnamefont {Assaf}}, \bibinfo {author}
      {\bibfnamefont {Don}\ \bibnamefont {Heiman}}, \bibinfo {author}
      {\bibfnamefont {Shou-Cheng}\ \bibnamefont {Zhang}}, \bibinfo {author}
      {\bibfnamefont {Chaoxing}\ \bibnamefont {Liu}}, \bibinfo {author}
      {\bibfnamefont {Moses H.~W.}\ \bibnamefont {Chan}}, \ and\ \bibinfo {author}
      {\bibfnamefont {Jagadeesh~S.}\ \bibnamefont {Moodera}},\ }\bibfield  {title}
      {\enquote {\bibinfo {title} {High-precision realization of robust quantum
      anomalous hall state in a hard ferromagnetic topological insulator},}\ }\href
      {\doibase 10.1038/nmat4204} {\bibfield  {journal} {\bibinfo  {journal}
      {Nature Mater.}\ }\textbf {\bibinfo {volume} {14}},\ \bibinfo {pages} {473}
      (\bibinfo {year} {2015})}\BibitemShut {NoStop}%
    \bibitem [{\citenamefont {Mogi}\ \emph {et~al.}(2015)\citenamefont {Mogi},
      \citenamefont {Yoshimi}, \citenamefont {Tsukazaki}, \citenamefont {Yasuda},
      \citenamefont {Kozuka}, \citenamefont {Takahashi}, \citenamefont {Kawasaki},\
      and\ \citenamefont {Tokura}}]{mogi2015}%
      \BibitemOpen
      \bibfield  {author} {\bibinfo {author} {\bibfnamefont {M.}~\bibnamefont
      {Mogi}}, \bibinfo {author} {\bibfnamefont {R.}~\bibnamefont {Yoshimi}},
      \bibinfo {author} {\bibfnamefont {A.}~\bibnamefont {Tsukazaki}}, \bibinfo
      {author} {\bibfnamefont {K.}~\bibnamefont {Yasuda}}, \bibinfo {author}
      {\bibfnamefont {Y.}~\bibnamefont {Kozuka}}, \bibinfo {author} {\bibfnamefont
      {K.~S.}\ \bibnamefont {Takahashi}}, \bibinfo {author} {\bibfnamefont
      {M.}~\bibnamefont {Kawasaki}}, \ and\ \bibinfo {author} {\bibfnamefont
      {Y.}~\bibnamefont {Tokura}},\ }\bibfield  {title} {\enquote {\bibinfo {title}
      {Magnetic modulation doping in topological insulators toward
      higher-temperature quantum anomalous hall effect},}\ }\href {\doibase
      10.1063/1.4935075} {\bibfield  {journal} {\bibinfo  {journal} {Appl. Phys.
      Lett.}\ }\textbf {\bibinfo {volume} {107}},\ \bibinfo {pages} {182401}
      (\bibinfo {year} {2015})}\BibitemShut {NoStop}%
    \bibitem [{\citenamefont {Watanabe}\ \emph {et~al.}(2019)\citenamefont
      {Watanabe}, \citenamefont {Yoshimi}, \citenamefont {Kawamura}, \citenamefont
      {Mogi}, \citenamefont {Tsukazaki}, \citenamefont {Yu}, \citenamefont
      {Nakajima}, \citenamefont {Takahashi}, \citenamefont {Kawasaki},\ and\
      \citenamefont {Tokura}}]{watanabe2019}%
      \BibitemOpen
      \bibfield  {author} {\bibinfo {author} {\bibfnamefont {R.}~\bibnamefont
      {Watanabe}}, \bibinfo {author} {\bibfnamefont {R.}~\bibnamefont {Yoshimi}},
      \bibinfo {author} {\bibfnamefont {M.}~\bibnamefont {Kawamura}}, \bibinfo
      {author} {\bibfnamefont {M.}~\bibnamefont {Mogi}}, \bibinfo {author}
      {\bibfnamefont {A.}~\bibnamefont {Tsukazaki}}, \bibinfo {author}
      {\bibfnamefont {X.~Z.}\ \bibnamefont {Yu}}, \bibinfo {author} {\bibfnamefont
      {K.}~\bibnamefont {Nakajima}}, \bibinfo {author} {\bibfnamefont {K.~S.}\
      \bibnamefont {Takahashi}}, \bibinfo {author} {\bibfnamefont {M.}~\bibnamefont
      {Kawasaki}}, \ and\ \bibinfo {author} {\bibfnamefont {Y.}~\bibnamefont
      {Tokura}},\ }\bibfield  {title} {\enquote {\bibinfo {title} {Quantum
      anomalous hall effect driven by magnetic proximity coupling in all-telluride
      based heterostructure},}\ }\href {\doibase 10.1063/1.5111891} {\bibfield
      {journal} {\bibinfo  {journal} {Appl. Phys. Lett.}\ }\textbf {\bibinfo
      {volume} {115}},\ \bibinfo {pages} {102403} (\bibinfo {year}
      {2019})}\BibitemShut {NoStop}%
    \bibitem [{\citenamefont {Deng}\ \emph {et~al.}(2020)\citenamefont {Deng},
      \citenamefont {Yu}, \citenamefont {Shi}, \citenamefont {Guo}, \citenamefont
      {Xu}, \citenamefont {Wang}, \citenamefont {Chen},\ and\ \citenamefont
      {Zhang}}]{deng2020}%
      \BibitemOpen
      \bibfield  {author} {\bibinfo {author} {\bibfnamefont {Yujun}\ \bibnamefont
      {Deng}}, \bibinfo {author} {\bibfnamefont {Yijun}\ \bibnamefont {Yu}},
      \bibinfo {author} {\bibfnamefont {Meng~Zhu}\ \bibnamefont {Shi}}, \bibinfo
      {author} {\bibfnamefont {Zhongxun}\ \bibnamefont {Guo}}, \bibinfo {author}
      {\bibfnamefont {Zihan}\ \bibnamefont {Xu}}, \bibinfo {author} {\bibfnamefont
      {Jing}\ \bibnamefont {Wang}}, \bibinfo {author} {\bibfnamefont {Xian~Hui}\
      \bibnamefont {Chen}}, \ and\ \bibinfo {author} {\bibfnamefont {Yuanbo}\
      \bibnamefont {Zhang}},\ }\bibfield  {title} {\enquote {\bibinfo {title}
      {Quantum anomalous hall effect in intrinsic magnetic topological insulator
      mnbi2te4},}\ }\href {\doibase 10.1126/science.aax8156} {\bibfield  {journal}
      {\bibinfo  {journal} {Science}\ }\textbf {\bibinfo {volume} {367}},\ \bibinfo
      {pages} {895--900} (\bibinfo {year} {2020})}\BibitemShut {NoStop}%
    \bibitem [{\citenamefont {Liu}\ \emph {et~al.}(2008)\citenamefont {Liu},
      \citenamefont {Qi}, \citenamefont {Dai}, \citenamefont {Fang},\ and\
      \citenamefont {Zhang}}]{liu2008}%
      \BibitemOpen
      \bibfield  {author} {\bibinfo {author} {\bibfnamefont {Chao-Xing}\
      \bibnamefont {Liu}}, \bibinfo {author} {\bibfnamefont {Xiao-Liang}\
      \bibnamefont {Qi}}, \bibinfo {author} {\bibfnamefont {Xi}~\bibnamefont
      {Dai}}, \bibinfo {author} {\bibfnamefont {Zhong}\ \bibnamefont {Fang}}, \
      and\ \bibinfo {author} {\bibfnamefont {Shou-Cheng}\ \bibnamefont {Zhang}},\
      }\bibfield  {title} {\enquote {\bibinfo {title} {Quantum anomalous hall
      effect in $\mathrm{Hg}_{1-y}\mathrm{Mn}_{y}\mathrm{Te}$ quantum wells},}\
      }\href {\doibase 10.1103/PhysRevLett.101.146802} {\bibfield  {journal}
      {\bibinfo  {journal} {Phys. Rev. Lett.}\ }\textbf {\bibinfo {volume} {101}},\
      \bibinfo {pages} {146802} (\bibinfo {year} {2008})}\BibitemShut {NoStop}%
    \bibitem [{\citenamefont {Yu}\ \emph {et~al.}(2010)\citenamefont {Yu},
      \citenamefont {Zhang}, \citenamefont {Zhang}, \citenamefont {Zhang},
      \citenamefont {Dai},\ and\ \citenamefont {Fang}}]{yu2010}%
      \BibitemOpen
      \bibfield  {author} {\bibinfo {author} {\bibfnamefont {Rui}\ \bibnamefont
      {Yu}}, \bibinfo {author} {\bibfnamefont {Wei}\ \bibnamefont {Zhang}},
      \bibinfo {author} {\bibfnamefont {Hai-Jun}\ \bibnamefont {Zhang}}, \bibinfo
      {author} {\bibfnamefont {Shou-Cheng}\ \bibnamefont {Zhang}}, \bibinfo
      {author} {\bibfnamefont {Xi}~\bibnamefont {Dai}}, \ and\ \bibinfo {author}
      {\bibfnamefont {Zhong}\ \bibnamefont {Fang}},\ }\bibfield  {title} {\enquote
      {\bibinfo {title} {{Quantized Anomalous Hall Effect in Magnetic Topological
      Insulators}},}\ }\href {\doibase 10.1126/science.1187485} {\bibfield
      {journal} {\bibinfo  {journal} {Science}\ }\textbf {\bibinfo {volume}
      {329}},\ \bibinfo {pages} {61--64} (\bibinfo {year} {2010})}\BibitemShut
      {NoStop}%
    \bibitem [{\citenamefont {Wang}\ \emph
      {et~al.}(2013{\natexlab{a}})\citenamefont {Wang}, \citenamefont {Lian},
      \citenamefont {Zhang}, \citenamefont {Xu},\ and\ \citenamefont
      {Zhang}}]{wang2013a}%
      \BibitemOpen
      \bibfield  {author} {\bibinfo {author} {\bibfnamefont {Jing}\ \bibnamefont
      {Wang}}, \bibinfo {author} {\bibfnamefont {Biao}\ \bibnamefont {Lian}},
      \bibinfo {author} {\bibfnamefont {Haijun}\ \bibnamefont {Zhang}}, \bibinfo
      {author} {\bibfnamefont {Yong}\ \bibnamefont {Xu}}, \ and\ \bibinfo {author}
      {\bibfnamefont {Shou-Cheng}\ \bibnamefont {Zhang}},\ }\bibfield  {title}
      {\enquote {\bibinfo {title} {Quantum anomalous hall effect with higher
      plateaus},}\ }\href {\doibase 10.1103/PhysRevLett.111.136801} {\bibfield
      {journal} {\bibinfo  {journal} {Phys. Rev. Lett.}\ }\textbf {\bibinfo
      {volume} {111}},\ \bibinfo {pages} {136801} (\bibinfo {year}
      {2013}{\natexlab{a}})}\BibitemShut {NoStop}%
    \bibitem [{\citenamefont {Essin}\ \emph {et~al.}(2009)\citenamefont {Essin},
      \citenamefont {Moore},\ and\ \citenamefont {Vanderbilt}}]{essin2009}%
      \BibitemOpen
      \bibfield  {author} {\bibinfo {author} {\bibfnamefont {Andrew~M.}\
      \bibnamefont {Essin}}, \bibinfo {author} {\bibfnamefont {Joel~E.}\
      \bibnamefont {Moore}}, \ and\ \bibinfo {author} {\bibfnamefont {David}\
      \bibnamefont {Vanderbilt}},\ }\bibfield  {title} {\enquote {\bibinfo {title}
      {Magnetoelectric polarizability and axion electrodynamics in crystalline
      insulators},}\ }\href {\doibase 10.1103/PhysRevLett.102.146805} {\bibfield
      {journal} {\bibinfo  {journal} {Phys. Rev. Lett.}\ }\textbf {\bibinfo
      {volume} {102}},\ \bibinfo {pages} {146805} (\bibinfo {year}
      {2009})}\BibitemShut {NoStop}%
    \bibitem [{\citenamefont {Coh}\ \emph {et~al.}(2011)\citenamefont {Coh},
      \citenamefont {Vanderbilt}, \citenamefont {Malashevich},\ and\ \citenamefont
      {Souza}}]{coh2011}%
      \BibitemOpen
      \bibfield  {author} {\bibinfo {author} {\bibfnamefont {Sinisa}\ \bibnamefont
      {Coh}}, \bibinfo {author} {\bibfnamefont {David}\ \bibnamefont {Vanderbilt}},
      \bibinfo {author} {\bibfnamefont {Andrei}\ \bibnamefont {Malashevich}}, \
      and\ \bibinfo {author} {\bibfnamefont {Ivo}\ \bibnamefont {Souza}},\
      }\bibfield  {title} {\enquote {\bibinfo {title} {Chern-simons orbital
      magnetoelectric coupling in generic insulators},}\ }\href {\doibase
      10.1103/PhysRevB.83.085108} {\bibfield  {journal} {\bibinfo  {journal} {Phys.
      Rev. B}\ }\textbf {\bibinfo {volume} {83}},\ \bibinfo {pages} {085108}
      (\bibinfo {year} {2011})}\BibitemShut {NoStop}%
    \bibitem [{\citenamefont {Wan}\ \emph {et~al.}(2012)\citenamefont {Wan},
      \citenamefont {Vishwanath},\ and\ \citenamefont {Savrasov}}]{wan2012}%
      \BibitemOpen
      \bibfield  {author} {\bibinfo {author} {\bibfnamefont {Xiangang}\
      \bibnamefont {Wan}}, \bibinfo {author} {\bibfnamefont {Ashvin}\ \bibnamefont
      {Vishwanath}}, \ and\ \bibinfo {author} {\bibfnamefont {Sergey~Y.}\
      \bibnamefont {Savrasov}},\ }\bibfield  {title} {\enquote {\bibinfo {title}
      {Computational design of axion insulators based on $5d$ spinel compounds},}\
      }\href {\doibase 10.1103/PhysRevLett.108.146601} {\bibfield  {journal}
      {\bibinfo  {journal} {Phys. Rev. Lett.}\ }\textbf {\bibinfo {volume} {108}},\
      \bibinfo {pages} {146601} (\bibinfo {year} {2012})}\BibitemShut {NoStop}%
    \bibitem [{\citenamefont {Turner}\ \emph {et~al.}(2012)\citenamefont {Turner},
      \citenamefont {Zhang}, \citenamefont {Mong},\ and\ \citenamefont
      {Vishwanath}}]{turner2012}%
      \BibitemOpen
      \bibfield  {author} {\bibinfo {author} {\bibfnamefont {Ari~M.}\ \bibnamefont
      {Turner}}, \bibinfo {author} {\bibfnamefont {Yi}~\bibnamefont {Zhang}},
      \bibinfo {author} {\bibfnamefont {Roger S.~K.}\ \bibnamefont {Mong}}, \ and\
      \bibinfo {author} {\bibfnamefont {Ashvin}\ \bibnamefont {Vishwanath}},\
      }\bibfield  {title} {\enquote {\bibinfo {title} {Quantized response and
      topology of magnetic insulators with inversion symmetry},}\ }\href {\doibase
      10.1103/PhysRevB.85.165120} {\bibfield  {journal} {\bibinfo  {journal} {Phys.
      Rev. B}\ }\textbf {\bibinfo {volume} {85}},\ \bibinfo {pages} {165120}
      (\bibinfo {year} {2012})}\BibitemShut {NoStop}%
    \bibitem [{\citenamefont {Varnava}\ and\ \citenamefont
      {Vanderbilt}(2018)}]{varnava2018}%
      \BibitemOpen
      \bibfield  {author} {\bibinfo {author} {\bibfnamefont {Nicodemos}\
      \bibnamefont {Varnava}}\ and\ \bibinfo {author} {\bibfnamefont {David}\
      \bibnamefont {Vanderbilt}},\ }\bibfield  {title} {\enquote {\bibinfo {title}
      {Surfaces of axion insulators},}\ }\href {\doibase
      10.1103/PhysRevB.98.245117} {\bibfield  {journal} {\bibinfo  {journal} {Phys.
      Rev. B}\ }\textbf {\bibinfo {volume} {98}},\ \bibinfo {pages} {245117}
      (\bibinfo {year} {2018})}\BibitemShut {NoStop}%
    \bibitem [{\citenamefont {Sekine}\ and\ \citenamefont
      {Nomura}(2021)}]{sekine2021}%
      \BibitemOpen
      \bibfield  {author} {\bibinfo {author} {\bibfnamefont {Akihiko}\ \bibnamefont
      {Sekine}}\ and\ \bibinfo {author} {\bibfnamefont {Kentaro}\ \bibnamefont
      {Nomura}},\ }\bibfield  {title} {\enquote {\bibinfo {title} {Axion
      electrodynamics in topological materials},}\ }\href {\doibase
      10.1063/5.0038804} {\bibfield  {journal} {\bibinfo  {journal} {J. Appl.
      Phys.}\ }\textbf {\bibinfo {volume} {129}},\ \bibinfo {pages} {141101}
      (\bibinfo {year} {2021})}\BibitemShut {NoStop}%
    \bibitem [{\citenamefont {Nomura}\ and\ \citenamefont
      {Nagaosa}(2011)}]{nomura2011}%
      \BibitemOpen
      \bibfield  {author} {\bibinfo {author} {\bibfnamefont {Kentaro}\ \bibnamefont
      {Nomura}}\ and\ \bibinfo {author} {\bibfnamefont {Naoto}\ \bibnamefont
      {Nagaosa}},\ }\bibfield  {title} {\enquote {\bibinfo {title}
      {Surface-quantized anomalous hall current and the magnetoelectric effect in
      magnetically disordered topological insulators},}\ }\href {\doibase
      10.1103/PhysRevLett.106.166802} {\bibfield  {journal} {\bibinfo  {journal}
      {Phys. Rev. Lett.}\ }\textbf {\bibinfo {volume} {106}},\ \bibinfo {pages}
      {166802} (\bibinfo {year} {2011})}\BibitemShut {NoStop}%
    \bibitem [{\citenamefont {Wang}\ \emph {et~al.}(2015)\citenamefont {Wang},
      \citenamefont {Lian}, \citenamefont {Qi},\ and\ \citenamefont
      {Zhang}}]{wang2015b}%
      \BibitemOpen
      \bibfield  {author} {\bibinfo {author} {\bibfnamefont {Jing}\ \bibnamefont
      {Wang}}, \bibinfo {author} {\bibfnamefont {Biao}\ \bibnamefont {Lian}},
      \bibinfo {author} {\bibfnamefont {Xiao-Liang}\ \bibnamefont {Qi}}, \ and\
      \bibinfo {author} {\bibfnamefont {Shou-Cheng}\ \bibnamefont {Zhang}},\
      }\bibfield  {title} {\enquote {\bibinfo {title} {{Quantized topological
      magnetoelectric effect of the zero-plateau quantum anomalous Hall state}},}\
      }\href {\doibase 10.1103/PhysRevB.92.081107} {\bibfield  {journal} {\bibinfo
      {journal} {Phys. Rev. B}\ }\textbf {\bibinfo {volume} {92}},\ \bibinfo
      {pages} {081107} (\bibinfo {year} {2015})}\BibitemShut {NoStop}%
    \bibitem [{\citenamefont {Morimoto}\ \emph {et~al.}(2015)\citenamefont
      {Morimoto}, \citenamefont {Furusaki},\ and\ \citenamefont
      {Nagaosa}}]{morimoto2015}%
      \BibitemOpen
      \bibfield  {author} {\bibinfo {author} {\bibfnamefont {Takahiro}\
      \bibnamefont {Morimoto}}, \bibinfo {author} {\bibfnamefont {Akira}\
      \bibnamefont {Furusaki}}, \ and\ \bibinfo {author} {\bibfnamefont {Naoto}\
      \bibnamefont {Nagaosa}},\ }\bibfield  {title} {\enquote {\bibinfo {title}
      {Topological magnetoelectric effects in thin films of topological
      insulators},}\ }\href {\doibase 10.1103/PhysRevB.92.085113} {\bibfield
      {journal} {\bibinfo  {journal} {Phys. Rev. B}\ }\textbf {\bibinfo {volume}
      {92}},\ \bibinfo {pages} {085113} (\bibinfo {year} {2015})}\BibitemShut
      {NoStop}%
    \bibitem [{\citenamefont {Wang}\ \emph {et~al.}(2014)\citenamefont {Wang},
      \citenamefont {Lian},\ and\ \citenamefont {Zhang}}]{wang2014a}%
      \BibitemOpen
      \bibfield  {author} {\bibinfo {author} {\bibfnamefont {Jing}\ \bibnamefont
      {Wang}}, \bibinfo {author} {\bibfnamefont {Biao}\ \bibnamefont {Lian}}, \
      and\ \bibinfo {author} {\bibfnamefont {Shou-Cheng}\ \bibnamefont {Zhang}},\
      }\bibfield  {title} {\enquote {\bibinfo {title} {Universal scaling of the
      quantum anomalous hall plateau transition},}\ }\href {\doibase
      10.1103/PhysRevB.89.085106} {\bibfield  {journal} {\bibinfo  {journal} {Phys.
      Rev. B}\ }\textbf {\bibinfo {volume} {89}},\ \bibinfo {pages} {085106}
      (\bibinfo {year} {2014})}\BibitemShut {NoStop}%
    \bibitem [{\citenamefont {Mogi}\ \emph
      {et~al.}(2017{\natexlab{a}})\citenamefont {Mogi}, \citenamefont {Kawamura},
      \citenamefont {Yoshimi}, \citenamefont {Tsukazaki}, \citenamefont {Kozuka},
      \citenamefont {Shirakawa}, \citenamefont {Takahashi}, \citenamefont
      {Kawasaki},\ and\ \citenamefont {Tokura}}]{mogi2017}%
      \BibitemOpen
      \bibfield  {author} {\bibinfo {author} {\bibfnamefont {M.}~\bibnamefont
      {Mogi}}, \bibinfo {author} {\bibfnamefont {M.}~\bibnamefont {Kawamura}},
      \bibinfo {author} {\bibfnamefont {R.}~\bibnamefont {Yoshimi}}, \bibinfo
      {author} {\bibfnamefont {A.}~\bibnamefont {Tsukazaki}}, \bibinfo {author}
      {\bibfnamefont {Y.}~\bibnamefont {Kozuka}}, \bibinfo {author} {\bibfnamefont
      {N.}~\bibnamefont {Shirakawa}}, \bibinfo {author} {\bibfnamefont {K.~S.}\
      \bibnamefont {Takahashi}}, \bibinfo {author} {\bibfnamefont {M.}~\bibnamefont
      {Kawasaki}}, \ and\ \bibinfo {author} {\bibfnamefont {Y.}~\bibnamefont
      {Tokura}},\ }\bibfield  {title} {\enquote {\bibinfo {title} {A magnetic
      heterostructure of topological insulators as a candidate for an axion
      insulator},}\ }\href {\doibase 10.1038/nmat4855} {\bibfield  {journal}
      {\bibinfo  {journal} {Nature Mater.}\ }\textbf {\bibinfo {volume} {16}},\
      \bibinfo {pages} {516--521} (\bibinfo {year}
      {2017}{\natexlab{a}})}\BibitemShut {NoStop}%
    \bibitem [{\citenamefont {Mogi}\ \emph
      {et~al.}(2017{\natexlab{b}})\citenamefont {Mogi}, \citenamefont {Kawamura},
      \citenamefont {Tsukazaki}, \citenamefont {Yoshimi}, \citenamefont
      {Takahashi}, \citenamefont {Kawasaki},\ and\ \citenamefont
      {Tokura}}]{mogi2017a}%
      \BibitemOpen
      \bibfield  {author} {\bibinfo {author} {\bibfnamefont {Masataka}\
      \bibnamefont {Mogi}}, \bibinfo {author} {\bibfnamefont {Minoru}\ \bibnamefont
      {Kawamura}}, \bibinfo {author} {\bibfnamefont {Atsushi}\ \bibnamefont
      {Tsukazaki}}, \bibinfo {author} {\bibfnamefont {Ryutaro}\ \bibnamefont
      {Yoshimi}}, \bibinfo {author} {\bibfnamefont {Kei~S.}\ \bibnamefont
      {Takahashi}}, \bibinfo {author} {\bibfnamefont {Masashi}\ \bibnamefont
      {Kawasaki}}, \ and\ \bibinfo {author} {\bibfnamefont {Yoshinori}\
      \bibnamefont {Tokura}},\ }\bibfield  {title} {\enquote {\bibinfo {title}
      {Tailoring tricolor structure of magnetic topological insulator for robust
      axion insulator},}\ }\href {\doibase 10.1126/sciadv.aao1669} {\bibfield
      {journal} {\bibinfo  {journal} {Sci. Adv.}\ }\textbf {\bibinfo {volume}
      {3}},\ \bibinfo {pages} {eaao1669} (\bibinfo {year}
      {2017}{\natexlab{b}})}\BibitemShut {NoStop}%
    \bibitem [{\citenamefont {Xiao}\ \emph {et~al.}(2018)\citenamefont {Xiao},
      \citenamefont {Jiang}, \citenamefont {Shin}, \citenamefont {Wang},
      \citenamefont {Wang}, \citenamefont {Zhao}, \citenamefont {Liu},
      \citenamefont {Wu}, \citenamefont {Chan}, \citenamefont {Samarth},\ and\
      \citenamefont {Chang}}]{xiao2018}%
      \BibitemOpen
      \bibfield  {author} {\bibinfo {author} {\bibfnamefont {Di}~\bibnamefont
      {Xiao}}, \bibinfo {author} {\bibfnamefont {Jue}\ \bibnamefont {Jiang}},
      \bibinfo {author} {\bibfnamefont {Jae-Ho}\ \bibnamefont {Shin}}, \bibinfo
      {author} {\bibfnamefont {Wenbo}\ \bibnamefont {Wang}}, \bibinfo {author}
      {\bibfnamefont {Fei}\ \bibnamefont {Wang}}, \bibinfo {author} {\bibfnamefont
      {Yi-Fan}\ \bibnamefont {Zhao}}, \bibinfo {author} {\bibfnamefont {Chaoxing}\
      \bibnamefont {Liu}}, \bibinfo {author} {\bibfnamefont {Weida}\ \bibnamefont
      {Wu}}, \bibinfo {author} {\bibfnamefont {Moses H.~W.}\ \bibnamefont {Chan}},
      \bibinfo {author} {\bibfnamefont {Nitin}\ \bibnamefont {Samarth}}, \ and\
      \bibinfo {author} {\bibfnamefont {Cui-Zu}\ \bibnamefont {Chang}},\ }\bibfield
       {title} {\enquote {\bibinfo {title} {{Realization of the Axion Insulator
      State in Quantum Anomalous Hall Sandwich Heterostructures}},}\ }\href
      {\doibase 10.1103/PhysRevLett.120.056801} {\bibfield  {journal} {\bibinfo
      {journal} {Phys. Rev. Lett.}\ }\textbf {\bibinfo {volume} {120}},\ \bibinfo
      {pages} {056801} (\bibinfo {year} {2018})}\BibitemShut {NoStop}%
    \bibitem [{\citenamefont {Grauer}\ \emph {et~al.}(2017)\citenamefont {Grauer},
      \citenamefont {Fijalkowski}, \citenamefont {Schreyeck}, \citenamefont
      {Winnerlein}, \citenamefont {Brunner}, \citenamefont {Thomale}, \citenamefont
      {Gould},\ and\ \citenamefont {Molenkamp}}]{grauer2017}%
      \BibitemOpen
      \bibfield  {author} {\bibinfo {author} {\bibfnamefont {S.}~\bibnamefont
      {Grauer}}, \bibinfo {author} {\bibfnamefont {K.~M.}\ \bibnamefont
      {Fijalkowski}}, \bibinfo {author} {\bibfnamefont {S.}~\bibnamefont
      {Schreyeck}}, \bibinfo {author} {\bibfnamefont {M.}~\bibnamefont
      {Winnerlein}}, \bibinfo {author} {\bibfnamefont {K.}~\bibnamefont {Brunner}},
      \bibinfo {author} {\bibfnamefont {R.}~\bibnamefont {Thomale}}, \bibinfo
      {author} {\bibfnamefont {C.}~\bibnamefont {Gould}}, \ and\ \bibinfo {author}
      {\bibfnamefont {L.~W.}\ \bibnamefont {Molenkamp}},\ }\bibfield  {title}
      {\enquote {\bibinfo {title} {Scaling of the quantum anomalous hall effect as
      an indicator of axion electrodynamics},}\ }\href {\doibase
      10.1103/PhysRevLett.118.246801} {\bibfield  {journal} {\bibinfo  {journal}
      {Phys. Rev. Lett.}\ }\textbf {\bibinfo {volume} {118}},\ \bibinfo {pages}
      {246801} (\bibinfo {year} {2017})}\BibitemShut {NoStop}%
    \bibitem [{\citenamefont {Liu}\ \emph {et~al.}(2020)\citenamefont {Liu},
      \citenamefont {Wang}, \citenamefont {Li}, \citenamefont {Wu}, \citenamefont
      {Li}, \citenamefont {Li}, \citenamefont {He}, \citenamefont {Xu},
      \citenamefont {Zhang},\ and\ \citenamefont {Wang}}]{liu2020}%
      \BibitemOpen
      \bibfield  {author} {\bibinfo {author} {\bibfnamefont {Chang}\ \bibnamefont
      {Liu}}, \bibinfo {author} {\bibfnamefont {Yongchao}\ \bibnamefont {Wang}},
      \bibinfo {author} {\bibfnamefont {Hao}\ \bibnamefont {Li}}, \bibinfo {author}
      {\bibfnamefont {Yang}\ \bibnamefont {Wu}}, \bibinfo {author} {\bibfnamefont
      {Yaoxin}\ \bibnamefont {Li}}, \bibinfo {author} {\bibfnamefont {Jiaheng}\
      \bibnamefont {Li}}, \bibinfo {author} {\bibfnamefont {Ke}~\bibnamefont {He}},
      \bibinfo {author} {\bibfnamefont {Yong}\ \bibnamefont {Xu}}, \bibinfo
      {author} {\bibfnamefont {Jinsong}\ \bibnamefont {Zhang}}, \ and\ \bibinfo
      {author} {\bibfnamefont {Yayu}\ \bibnamefont {Wang}},\ }\bibfield  {title}
      {\enquote {\bibinfo {title} {Robust axion insulator and chern insulator
      phases in a two-dimensional antiferromagnetic topological insulator},}\
      }\href {\doibase 10.1038/s41563-019-0573-3} {\bibfield  {journal} {\bibinfo
      {journal} {Nature Mat.}\ }\textbf {\bibinfo {volume} {19}},\ \bibinfo {pages}
      {522--527} (\bibinfo {year} {2020})}\BibitemShut {NoStop}%
    \bibitem [{\citenamefont {Zhang}\ \emph
      {et~al.}(2019{\natexlab{a}})\citenamefont {Zhang}, \citenamefont {Shi},
      \citenamefont {Zhu}, \citenamefont {Xing}, \citenamefont {Zhang},\ and\
      \citenamefont {Wang}}]{zhang2019}%
      \BibitemOpen
      \bibfield  {author} {\bibinfo {author} {\bibfnamefont {Dongqin}\ \bibnamefont
      {Zhang}}, \bibinfo {author} {\bibfnamefont {Minji}\ \bibnamefont {Shi}},
      \bibinfo {author} {\bibfnamefont {Tongshuai}\ \bibnamefont {Zhu}}, \bibinfo
      {author} {\bibfnamefont {Dingyu}\ \bibnamefont {Xing}}, \bibinfo {author}
      {\bibfnamefont {Haijun}\ \bibnamefont {Zhang}}, \ and\ \bibinfo {author}
      {\bibfnamefont {Jing}\ \bibnamefont {Wang}},\ }\bibfield  {title} {\enquote
      {\bibinfo {title} {Topological axion states in the magnetic insulator
      ${\mathrm{mnbi}}_{2}{\mathrm{te}}_{4}$ with the quantized magnetoelectric
      effect},}\ }\href {\doibase 10.1103/PhysRevLett.122.206401} {\bibfield
      {journal} {\bibinfo  {journal} {Phys. Rev. Lett.}\ }\textbf {\bibinfo
      {volume} {122}},\ \bibinfo {pages} {206401} (\bibinfo {year}
      {2019}{\natexlab{a}})}\BibitemShut {NoStop}%
    \bibitem [{\citenamefont {Li}\ \emph {et~al.}(2019)\citenamefont {Li},
      \citenamefont {Li}, \citenamefont {Du}, \citenamefont {Wang}, \citenamefont
      {Gu}, \citenamefont {Zhang}, \citenamefont {He}, \citenamefont {Duan},\ and\
      \citenamefont {Xu}}]{li2019}%
      \BibitemOpen
      \bibfield  {author} {\bibinfo {author} {\bibfnamefont {Jiaheng}\ \bibnamefont
      {Li}}, \bibinfo {author} {\bibfnamefont {Yang}\ \bibnamefont {Li}}, \bibinfo
      {author} {\bibfnamefont {Shiqiao}\ \bibnamefont {Du}}, \bibinfo {author}
      {\bibfnamefont {Zun}\ \bibnamefont {Wang}}, \bibinfo {author} {\bibfnamefont
      {Bing-Lin}\ \bibnamefont {Gu}}, \bibinfo {author} {\bibfnamefont
      {Shou-Cheng}\ \bibnamefont {Zhang}}, \bibinfo {author} {\bibfnamefont
      {Ke}~\bibnamefont {He}}, \bibinfo {author} {\bibfnamefont {Wenhui}\
      \bibnamefont {Duan}}, \ and\ \bibinfo {author} {\bibfnamefont {Yong}\
      \bibnamefont {Xu}},\ }\bibfield  {title} {\enquote {\bibinfo {title}
      {Intrinsic magnetic topological insulators in van der waals layered
      mnbi2te4-family materials},}\ }\href {\doibase 10.1126/sciadv.aaw5685}
      {\bibfield  {journal} {\bibinfo  {journal} {Sci. Adv.}\ }\textbf {\bibinfo
      {volume} {5}},\ \bibinfo {pages} {eaaw5685} (\bibinfo {year}
      {2019})}\BibitemShut {NoStop}%
    \bibitem [{\citenamefont {Otrokov}\ \emph {et~al.}(2019)\citenamefont
      {Otrokov}, \citenamefont {Rusinov}, \citenamefont {Blanco-Rey}, \citenamefont
      {Hoffmann}, \citenamefont {Vyazovskaya}, \citenamefont {Eremeev},
      \citenamefont {Ernst}, \citenamefont {Echenique}, \citenamefont {Arnau},\
      and\ \citenamefont {Chulkov}}]{otrokov2019a}%
      \BibitemOpen
      \bibfield  {author} {\bibinfo {author} {\bibfnamefont {M.~M.}\ \bibnamefont
      {Otrokov}}, \bibinfo {author} {\bibfnamefont {I.~P.}\ \bibnamefont
      {Rusinov}}, \bibinfo {author} {\bibfnamefont {M.}~\bibnamefont {Blanco-Rey}},
      \bibinfo {author} {\bibfnamefont {M.}~\bibnamefont {Hoffmann}}, \bibinfo
      {author} {\bibfnamefont {A.~Yu.}\ \bibnamefont {Vyazovskaya}}, \bibinfo
      {author} {\bibfnamefont {S.~V.}\ \bibnamefont {Eremeev}}, \bibinfo {author}
      {\bibfnamefont {A.}~\bibnamefont {Ernst}}, \bibinfo {author} {\bibfnamefont
      {P.~M.}\ \bibnamefont {Echenique}}, \bibinfo {author} {\bibfnamefont
      {A.}~\bibnamefont {Arnau}}, \ and\ \bibinfo {author} {\bibfnamefont {E.~V.}\
      \bibnamefont {Chulkov}},\ }\bibfield  {title} {\enquote {\bibinfo {title}
      {Unique thickness-dependent properties of the van der waals interlayer
      antiferromagnet ${\mathrm{mnbi}}_{2}{\mathrm{te}}_{4}$ films},}\ }\href
      {\doibase 10.1103/PhysRevLett.122.107202} {\bibfield  {journal} {\bibinfo
      {journal} {Phys. Rev. Lett.}\ }\textbf {\bibinfo {volume} {122}},\ \bibinfo
      {pages} {107202} (\bibinfo {year} {2019})}\BibitemShut {NoStop}%
    \bibitem [{\citenamefont {Liu}\ and\ \citenamefont {Wang}(2020)}]{liuzc2020a}%
      \BibitemOpen
      \bibfield  {author} {\bibinfo {author} {\bibfnamefont {Zhaochen}\
      \bibnamefont {Liu}}\ and\ \bibinfo {author} {\bibfnamefont {Jing}\
      \bibnamefont {Wang}},\ }\bibfield  {title} {\enquote {\bibinfo {title}
      {Anisotropic topological magnetoelectric effect in axion insulators},}\
      }\href {\doibase 10.1103/PhysRevB.101.205130} {\bibfield  {journal} {\bibinfo
       {journal} {Phys. Rev. B}\ }\textbf {\bibinfo {volume} {101}},\ \bibinfo
      {pages} {205130} (\bibinfo {year} {2020})}\BibitemShut {NoStop}%
    \bibitem [{\citenamefont {Allen}\ \emph {et~al.}(2019)\citenamefont {Allen},
      \citenamefont {Cui}, \citenamefont {Yue~Ma}, \citenamefont {Mogi},
      \citenamefont {Kawamura}, \citenamefont {Fulga}, \citenamefont
      {Goldhaber-Gordon}, \citenamefont {Tokura},\ and\ \citenamefont
      {Shen}}]{allen2019}%
      \BibitemOpen
      \bibfield  {author} {\bibinfo {author} {\bibfnamefont {Monica}\ \bibnamefont
      {Allen}}, \bibinfo {author} {\bibfnamefont {Yongtao}\ \bibnamefont {Cui}},
      \bibinfo {author} {\bibfnamefont {Eric}\ \bibnamefont {Yue~Ma}}, \bibinfo
      {author} {\bibfnamefont {Masataka}\ \bibnamefont {Mogi}}, \bibinfo {author}
      {\bibfnamefont {Minoru}\ \bibnamefont {Kawamura}}, \bibinfo {author}
      {\bibfnamefont {Ion~Cosma}\ \bibnamefont {Fulga}}, \bibinfo {author}
      {\bibfnamefont {David}\ \bibnamefont {Goldhaber-Gordon}}, \bibinfo {author}
      {\bibfnamefont {Yoshinori}\ \bibnamefont {Tokura}}, \ and\ \bibinfo {author}
      {\bibfnamefont {Zhi-Xun}\ \bibnamefont {Shen}},\ }\bibfield  {title}
      {\enquote {\bibinfo {title} {Visualization of an axion insulating state at
      the transition between 2 chiral quantum anomalous hall states},}\ }\href
      {\doibase 10.1073/pnas.1818255116} {\bibfield  {journal} {\bibinfo  {journal}
      {Pro. Natl. Acad. Sci.}\ }\textbf {\bibinfo {volume} {116}},\ \bibinfo
      {pages} {14511--14515} (\bibinfo {year} {2019})}\BibitemShut {NoStop}%
    \bibitem [{\citenamefont {Li}\ \emph {et~al.}()\citenamefont {Li},
      \citenamefont {Liu}, \citenamefont {Wang}, \citenamefont {Lian},
      \citenamefont {Li}, \citenamefont {Wu}, \citenamefont {Zhang},\ and\
      \citenamefont {Wang}}]{li2021}%
      \BibitemOpen
      \bibfield  {author} {\bibinfo {author} {\bibfnamefont {Yaoxin}\ \bibnamefont
      {Li}}, \bibinfo {author} {\bibfnamefont {Chang}\ \bibnamefont {Liu}},
      \bibinfo {author} {\bibfnamefont {Yongchao}\ \bibnamefont {Wang}}, \bibinfo
      {author} {\bibfnamefont {Zichen}\ \bibnamefont {Lian}}, \bibinfo {author}
      {\bibfnamefont {Hao}\ \bibnamefont {Li}}, \bibinfo {author} {\bibfnamefont
      {Yang}\ \bibnamefont {Wu}}, \bibinfo {author} {\bibfnamefont {Jinsong}\
      \bibnamefont {Zhang}}, \ and\ \bibinfo {author} {\bibfnamefont {Yayu}\
      \bibnamefont {Wang}},\ }\bibfield  {title} {\enquote {\bibinfo {title}
      {Nonlocal transport in axion insulator state of mnbi2te4},}\ }\href@noop {}
      {\bibinfo  {journal} {arXiv: 2105.10390}\ }\BibitemShut {NoStop}%
    \bibitem [{\citenamefont {Lin}\ \emph {et~al.}()\citenamefont {Lin},
      \citenamefont {Feng}, \citenamefont {Wang}, \citenamefont {Lian},
      \citenamefont {Li}, \citenamefont {Wu}, \citenamefont {Liu}, \citenamefont
      {Wang}, \citenamefont {Zhang}, \citenamefont {Wang}, \citenamefont {Zhou},\
      and\ \citenamefont {Shen}}]{lin2021}%
      \BibitemOpen
    \bibfield  {journal} {  }\bibfield  {author} {\bibinfo {author} {\bibfnamefont
      {Weiyan}\ \bibnamefont {Lin}}, \bibinfo {author} {\bibfnamefont {Yang}\
      \bibnamefont {Feng}}, \bibinfo {author} {\bibfnamefont {Yongchao}\
      \bibnamefont {Wang}}, \bibinfo {author} {\bibfnamefont {Zichen}\ \bibnamefont
      {Lian}}, \bibinfo {author} {\bibfnamefont {Hao}\ \bibnamefont {Li}}, \bibinfo
      {author} {\bibfnamefont {Yang}\ \bibnamefont {Wu}}, \bibinfo {author}
      {\bibfnamefont {Chang}\ \bibnamefont {Liu}}, \bibinfo {author} {\bibfnamefont
      {Yihua}\ \bibnamefont {Wang}}, \bibinfo {author} {\bibfnamefont {Jinsong}\
      \bibnamefont {Zhang}}, \bibinfo {author} {\bibfnamefont {Yayu}\ \bibnamefont
      {Wang}}, \bibinfo {author} {\bibfnamefont {Xiaodong}\ \bibnamefont {Zhou}}, \
      and\ \bibinfo {author} {\bibfnamefont {Jian}\ \bibnamefont {Shen}},\
      }\bibfield  {title} {\enquote {\bibinfo {title} {Direct visualization of edge
      state in even-layer mnbi2te4 at zero magnetic field},}\ }\href@noop {}
      {\bibinfo  {journal} {arXiv:2105.10234}\ }\BibitemShut {NoStop}%
    \bibitem [{\citenamefont {Mong}\ \emph {et~al.}(2010)\citenamefont {Mong},
      \citenamefont {Essin},\ and\ \citenamefont {Moore}}]{mong2010}%
      \BibitemOpen
    \bibfield  {journal} {  }\bibfield  {author} {\bibinfo {author} {\bibfnamefont
      {Roger S.~K.}\ \bibnamefont {Mong}}, \bibinfo {author} {\bibfnamefont
      {Andrew~M.}\ \bibnamefont {Essin}}, \ and\ \bibinfo {author} {\bibfnamefont
      {Joel~E.}\ \bibnamefont {Moore}},\ }\bibfield  {title} {\enquote {\bibinfo
      {title} {Antiferromagnetic topological insulators},}\ }\href {\doibase
      10.1103/PhysRevB.81.245209} {\bibfield  {journal} {\bibinfo  {journal} {Phys.
      Rev. B}\ }\textbf {\bibinfo {volume} {81}},\ \bibinfo {pages} {245209}
      (\bibinfo {year} {2010})}\BibitemShut {NoStop}%
    \bibitem [{\citenamefont {Liu}\ \emph {et~al.}(2010)\citenamefont {Liu},
      \citenamefont {Zhang}, \citenamefont {Yan}, \citenamefont {Qi}, \citenamefont
      {Frauenheim}, \citenamefont {Dai}, \citenamefont {Fang},\ and\ \citenamefont
      {Zhang}}]{liu2010a}%
      \BibitemOpen
      \bibfield  {author} {\bibinfo {author} {\bibfnamefont {Chao-Xing}\
      \bibnamefont {Liu}}, \bibinfo {author} {\bibfnamefont {Hai-Jun}\ \bibnamefont
      {Zhang}}, \bibinfo {author} {\bibfnamefont {Binghai}\ \bibnamefont {Yan}},
      \bibinfo {author} {\bibfnamefont {Xiao-Liang}\ \bibnamefont {Qi}}, \bibinfo
      {author} {\bibfnamefont {Thomas}\ \bibnamefont {Frauenheim}}, \bibinfo
      {author} {\bibfnamefont {Xi}~\bibnamefont {Dai}}, \bibinfo {author}
      {\bibfnamefont {Zhong}\ \bibnamefont {Fang}}, \ and\ \bibinfo {author}
      {\bibfnamefont {Shou-Cheng}\ \bibnamefont {Zhang}},\ }\bibfield  {title}
      {\enquote {\bibinfo {title} {Oscillatory crossover from two-dimensional to
      three-dimensional topological insulators},}\ }\href {\doibase
      10.1103/PhysRevB.81.041307} {\bibfield  {journal} {\bibinfo  {journal} {Phys.
      Rev. B}\ }\textbf {\bibinfo {volume} {81}},\ \bibinfo {pages} {041307}
      (\bibinfo {year} {2010})}\BibitemShut {NoStop}%
    \bibitem [{\citenamefont {Lu}\ \emph {et~al.}(2010)\citenamefont {Lu},
      \citenamefont {Shan}, \citenamefont {Yao}, \citenamefont {Niu},\ and\
      \citenamefont {Shen}}]{lu2010}%
      \BibitemOpen
      \bibfield  {author} {\bibinfo {author} {\bibfnamefont {Hai-Zhou}\
      \bibnamefont {Lu}}, \bibinfo {author} {\bibfnamefont {Wen-Yu}\ \bibnamefont
      {Shan}}, \bibinfo {author} {\bibfnamefont {Wang}\ \bibnamefont {Yao}},
      \bibinfo {author} {\bibfnamefont {Qian}\ \bibnamefont {Niu}}, \ and\ \bibinfo
      {author} {\bibfnamefont {Shun-Qing}\ \bibnamefont {Shen}},\ }\bibfield
      {title} {\enquote {\bibinfo {title} {Massive dirac fermions and spin physics
      in an ultrathin film of topological insulator},}\ }\href {\doibase
      10.1103/PhysRevB.81.115407} {\bibfield  {journal} {\bibinfo  {journal} {Phys.
      Rev. B}\ }\textbf {\bibinfo {volume} {81}},\ \bibinfo {pages} {115407}
      (\bibinfo {year} {2010})}\BibitemShut {NoStop}%
    \bibitem [{\citenamefont {Zhang}\ \emph
      {et~al.}(2019{\natexlab{b}})\citenamefont {Zhang}, \citenamefont {Liu},\ and\
      \citenamefont {Wang}}]{zhangjl2019}%
      \BibitemOpen
      \bibfield  {author} {\bibinfo {author} {\bibfnamefont {Jinlong}\ \bibnamefont
      {Zhang}}, \bibinfo {author} {\bibfnamefont {Zhaochen}\ \bibnamefont {Liu}}, \
      and\ \bibinfo {author} {\bibfnamefont {Jing}\ \bibnamefont {Wang}},\
      }\bibfield  {title} {\enquote {\bibinfo {title} {In-plane
      magnetic-field-induced quantum anomalous hall plateau transition},}\ }\href
      {\doibase 10.1103/PhysRevB.100.165117} {\bibfield  {journal} {\bibinfo
      {journal} {Phys. Rev. B}\ }\textbf {\bibinfo {volume} {100}},\ \bibinfo
      {pages} {165117} (\bibinfo {year} {2019}{\natexlab{b}})}\BibitemShut
      {NoStop}%
    \bibitem [{\citenamefont {Sun}\ \emph {et~al.}(2020)\citenamefont {Sun},
      \citenamefont {Wang}, \citenamefont {Zhang}, \citenamefont {Chen},
      \citenamefont {Zhao}, \citenamefont {Liu}, \citenamefont {Liu}, \citenamefont
      {Chen}, \citenamefont {Lu},\ and\ \citenamefont {Xie}}]{sun2020}%
      \BibitemOpen
      \bibfield  {author} {\bibinfo {author} {\bibfnamefont {Hai-Peng}\
      \bibnamefont {Sun}}, \bibinfo {author} {\bibfnamefont {C.~M.}\ \bibnamefont
      {Wang}}, \bibinfo {author} {\bibfnamefont {Song-Bo}\ \bibnamefont {Zhang}},
      \bibinfo {author} {\bibfnamefont {Rui}\ \bibnamefont {Chen}}, \bibinfo
      {author} {\bibfnamefont {Yue}\ \bibnamefont {Zhao}}, \bibinfo {author}
      {\bibfnamefont {Chang}\ \bibnamefont {Liu}}, \bibinfo {author} {\bibfnamefont
      {Qihang}\ \bibnamefont {Liu}}, \bibinfo {author} {\bibfnamefont {Chaoyu}\
      \bibnamefont {Chen}}, \bibinfo {author} {\bibfnamefont {Hai-Zhou}\
      \bibnamefont {Lu}}, \ and\ \bibinfo {author} {\bibfnamefont {X.~C.}\
      \bibnamefont {Xie}},\ }\bibfield  {title} {\enquote {\bibinfo {title}
      {Analytical solution for the surface states of the antiferromagnetic
      topological insulator ${\mathrm{mnbi}}_{2}{\mathrm{te}}_{4}$},}\ }\href
      {\doibase 10.1103/PhysRevB.102.241406} {\bibfield  {journal} {\bibinfo
      {journal} {Phys. Rev. B}\ }\textbf {\bibinfo {volume} {102}},\ \bibinfo
      {pages} {241406} (\bibinfo {year} {2020})}\BibitemShut {NoStop}%
    \bibitem [{\citenamefont {Bernevig}\ \emph {et~al.}(2006)\citenamefont
      {Bernevig}, \citenamefont {Hughes},\ and\ \citenamefont
      {Zhang}}]{bernevig2006c}%
      \BibitemOpen
      \bibfield  {author} {\bibinfo {author} {\bibfnamefont {B.~Andrei}\
      \bibnamefont {Bernevig}}, \bibinfo {author} {\bibfnamefont {Taylor~L.}\
      \bibnamefont {Hughes}}, \ and\ \bibinfo {author} {\bibfnamefont {Shou-Cheng}\
      \bibnamefont {Zhang}},\ }\bibfield  {title} {\enquote {\bibinfo {title}
      {Quantum spin hall effect and topological phase transition in hgte quantum
      wells},}\ }\href {\doibase 10.1126/science.1133734} {\bibfield  {journal}
      {\bibinfo  {journal} {Science}\ }\textbf {\bibinfo {volume} {314}},\ \bibinfo
      {pages} {1757--1761} (\bibinfo {year} {2006})}\BibitemShut {NoStop}%
    \bibitem [{sup()}]{supple}%
      \BibitemOpen
      \href@noop {} {}\bibinfo {note} {See Supplemental Material for technical
      details.}\BibitemShut {Stop}%
    \bibitem [{\citenamefont {Altland}\ and\ \citenamefont
      {Simons}(2010)}]{altland2010}%
      \BibitemOpen
      \bibfield  {author} {\bibinfo {author} {\bibfnamefont {Alexander}\
      \bibnamefont {Altland}}\ and\ \bibinfo {author} {\bibfnamefont {Ben}\
      \bibnamefont {Simons}},\ }\href@noop {} {\emph {\bibinfo {title} {Condensed
      matter field theory}}}\ (\bibinfo  {publisher} {Cambridge university press},\
      \bibinfo {year} {2010})\BibitemShut {NoStop}%
    \bibitem [{\citenamefont {Groth}\ \emph {et~al.}(2009)\citenamefont {Groth},
      \citenamefont {Wimmer}, \citenamefont {Akhmerov}, \citenamefont
      {Tworzyd\l{}o},\ and\ \citenamefont {Beenakker}}]{groth2009}%
      \BibitemOpen
      \bibfield  {author} {\bibinfo {author} {\bibfnamefont {C.~W.}\ \bibnamefont
      {Groth}}, \bibinfo {author} {\bibfnamefont {M.}~\bibnamefont {Wimmer}},
      \bibinfo {author} {\bibfnamefont {A.~R.}\ \bibnamefont {Akhmerov}}, \bibinfo
      {author} {\bibfnamefont {J.}~\bibnamefont {Tworzyd\l{}o}}, \ and\ \bibinfo
      {author} {\bibfnamefont {C.~W.~J.}\ \bibnamefont {Beenakker}},\ }\bibfield
      {title} {\enquote {\bibinfo {title} {Theory of the topological anderson
      insulator},}\ }\href {\doibase 10.1103/PhysRevLett.103.196805} {\bibfield
      {journal} {\bibinfo  {journal} {Phys. Rev. Lett.}\ }\textbf {\bibinfo
      {volume} {103}},\ \bibinfo {pages} {196805} (\bibinfo {year}
      {2009})}\BibitemShut {NoStop}%
    \bibitem [{\citenamefont {Li}\ \emph {et~al.}(2009)\citenamefont {Li},
      \citenamefont {Chu}, \citenamefont {Jain},\ and\ \citenamefont
      {Shen}}]{li2009}%
      \BibitemOpen
      \bibfield  {author} {\bibinfo {author} {\bibfnamefont {Jian}\ \bibnamefont
      {Li}}, \bibinfo {author} {\bibfnamefont {Rui-Lin}\ \bibnamefont {Chu}},
      \bibinfo {author} {\bibfnamefont {J.~K.}\ \bibnamefont {Jain}}, \ and\
      \bibinfo {author} {\bibfnamefont {Shun-Qing}\ \bibnamefont {Shen}},\
      }\bibfield  {title} {\enquote {\bibinfo {title} {Topological anderson
      insulator},}\ }\href {\doibase 10.1103/PhysRevLett.102.136806} {\bibfield
      {journal} {\bibinfo  {journal} {Phys. Rev. Lett.}\ }\textbf {\bibinfo
      {volume} {102}},\ \bibinfo {pages} {136806} (\bibinfo {year}
      {2009})}\BibitemShut {NoStop}%
    \bibitem [{\citenamefont {Jiang}\ \emph {et~al.}(2009)\citenamefont {Jiang},
      \citenamefont {Wang}, \citenamefont {Sun},\ and\ \citenamefont
      {Xie}}]{jiang2009}%
      \BibitemOpen
      \bibfield  {author} {\bibinfo {author} {\bibfnamefont {Hua}\ \bibnamefont
      {Jiang}}, \bibinfo {author} {\bibfnamefont {Lei}\ \bibnamefont {Wang}},
      \bibinfo {author} {\bibfnamefont {Qing-feng}\ \bibnamefont {Sun}}, \ and\
      \bibinfo {author} {\bibfnamefont {X.~C.}\ \bibnamefont {Xie}},\ }\bibfield
      {title} {\enquote {\bibinfo {title} {Numerical study of the topological
      anderson insulator in hgte/cdte quantum wells},}\ }\href {\doibase
      10.1103/PhysRevB.80.165316} {\bibfield  {journal} {\bibinfo  {journal} {Phys.
      Rev. B}\ }\textbf {\bibinfo {volume} {80}},\ \bibinfo {pages} {165316}
      (\bibinfo {year} {2009})}\BibitemShut {NoStop}%
    \bibitem [{\citenamefont {Kane}\ and\ \citenamefont {Fisher}(2007)}]{kane2007}%
      \BibitemOpen
      \bibfield  {author} {\bibinfo {author} {\bibfnamefont {C.~L.}\ \bibnamefont
      {Kane}}\ and\ \bibinfo {author} {\bibfnamefont {Matthew P.~A.}\ \bibnamefont
      {Fisher}},\ }\enquote {\bibinfo {title} {Edge-state transport},}\ in\ \href
      {\doibase 10.1002/9783527617258.ch4} {\emph {\bibinfo {booktitle}
      {Perspectives in Quantum Hall Effects}}}\ (\bibinfo  {publisher} {Wiley-VCH
      Verlag GmbH, New York},\ \bibinfo {year} {2007})\ pp.\ \bibinfo {pages}
      {109--159}\BibitemShut {NoStop}%
    \bibitem [{\citenamefont {Giamarchi}(2003)}]{giamarchi2003}%
      \BibitemOpen
      \bibfield  {author} {\bibinfo {author} {\bibfnamefont {T.}~\bibnamefont
      {Giamarchi}},\ }\href@noop {} {\emph {\bibinfo {title} {Quantum physics in
      one dimension}}},\ Vol.\ \bibinfo {volume} {121}\ (\bibinfo  {publisher}
      {Clarendon press},\ \bibinfo {year} {2003})\BibitemShut {NoStop}%
    \bibitem [{\citenamefont {Groth}\ \emph {et~al.}(2014)\citenamefont {Groth},
      \citenamefont {Wimmer}, \citenamefont {Akhmerov},\ and\ \citenamefont
      {Waintal}}]{groth2014}%
      \BibitemOpen
      \bibfield  {author} {\bibinfo {author} {\bibfnamefont {C.~W.}\ \bibnamefont
      {Groth}}, \bibinfo {author} {\bibfnamefont {M.}~\bibnamefont {Wimmer}},
      \bibinfo {author} {\bibfnamefont {A.~R.}\ \bibnamefont {Akhmerov}}, \ and\
      \bibinfo {author} {\bibfnamefont {X.}~\bibnamefont {Waintal}},\ }\bibfield
      {title} {\enquote {\bibinfo {title} {Kwant: a software package for quantum
      transport},}\ }\href {\doibase 10.1088/1367-2630/16/6/063065} {\bibfield
      {journal} {\bibinfo  {journal} {New J. Phys.}\ }\textbf {\bibinfo {volume}
      {16}},\ \bibinfo {pages} {063065} (\bibinfo {year} {2014})}\BibitemShut
      {NoStop}%
    \bibitem [{\citenamefont {B\"uttiker}(1986)}]{buttiker1986}%
      \BibitemOpen
      \bibfield  {author} {\bibinfo {author} {\bibfnamefont {M.}~\bibnamefont
      {B\"uttiker}},\ }\bibfield  {title} {\enquote {\bibinfo {title}
      {Four-terminal phase-coherent conductance},}\ }\href {\doibase
      10.1103/PhysRevLett.57.1761} {\bibfield  {journal} {\bibinfo  {journal}
      {Phys. Rev. Lett.}\ }\textbf {\bibinfo {volume} {57}},\ \bibinfo {pages}
      {1761--1764} (\bibinfo {year} {1986})}\BibitemShut {NoStop}%
    \bibitem [{\citenamefont {B\"uttiker}(1988)}]{buttiker1988}%
      \BibitemOpen
      \bibfield  {author} {\bibinfo {author} {\bibfnamefont {M.}~\bibnamefont
      {B\"uttiker}},\ }\bibfield  {title} {\enquote {\bibinfo {title} {Absence of
      backscattering in the quantum hall effect in multiprobe conductors},}\ }\href
      {\doibase 10.1103/PhysRevB.38.9375} {\bibfield  {journal} {\bibinfo
      {journal} {Phys. Rev. B}\ }\textbf {\bibinfo {volume} {38}},\ \bibinfo
      {pages} {9375--9389} (\bibinfo {year} {1988})}\BibitemShut {NoStop}%
    \bibitem [{\citenamefont {Wang}\ \emph
      {et~al.}(2013{\natexlab{b}})\citenamefont {Wang}, \citenamefont {Lian},
      \citenamefont {Zhang},\ and\ \citenamefont {Zhang}}]{wang2013b}%
      \BibitemOpen
      \bibfield  {author} {\bibinfo {author} {\bibfnamefont {Jing}\ \bibnamefont
      {Wang}}, \bibinfo {author} {\bibfnamefont {Biao}\ \bibnamefont {Lian}},
      \bibinfo {author} {\bibfnamefont {Haijun}\ \bibnamefont {Zhang}}, \ and\
      \bibinfo {author} {\bibfnamefont {Shou-Cheng}\ \bibnamefont {Zhang}},\
      }\bibfield  {title} {\enquote {\bibinfo {title} {Anomalous edge transport in
      the quantum anomalous hall state},}\ }\href {\doibase
      10.1103/PhysRevLett.111.086803} {\bibfield  {journal} {\bibinfo  {journal}
      {Phys. Rev. Lett.}\ }\textbf {\bibinfo {volume} {111}},\ \bibinfo {pages}
      {086803} (\bibinfo {year} {2013}{\natexlab{b}})}\BibitemShut {NoStop}%
    \bibitem [{\citenamefont {Roth}\ \emph {et~al.}(2009)\citenamefont {Roth},
      \citenamefont {Br{\"u}ne}, \citenamefont {Buhmann}, \citenamefont
      {Molenkamp}, \citenamefont {Maciejko}, \citenamefont {Qi},\ and\
      \citenamefont {Zhang}}]{roth2009}%
      \BibitemOpen
      \bibfield  {author} {\bibinfo {author} {\bibfnamefont {Andreas}\ \bibnamefont
      {Roth}}, \bibinfo {author} {\bibfnamefont {Christoph}\ \bibnamefont
      {Br{\"u}ne}}, \bibinfo {author} {\bibfnamefont {Hartmut}\ \bibnamefont
      {Buhmann}}, \bibinfo {author} {\bibfnamefont {Laurens~W.}\ \bibnamefont
      {Molenkamp}}, \bibinfo {author} {\bibfnamefont {Joseph}\ \bibnamefont
      {Maciejko}}, \bibinfo {author} {\bibfnamefont {Xiao-Liang}\ \bibnamefont
      {Qi}}, \ and\ \bibinfo {author} {\bibfnamefont {Shou-Cheng}\ \bibnamefont
      {Zhang}},\ }\bibfield  {title} {\enquote {\bibinfo {title} {Nonlocal
      transport in the quantum spin hall state},}\ }\href {\doibase
      10.1126/science.1174736} {\bibfield  {journal} {\bibinfo  {journal}
      {Science}\ }\textbf {\bibinfo {volume} {325}},\ \bibinfo {pages} {294--297}
      (\bibinfo {year} {2009})}\BibitemShut {NoStop}%
    \bibitem [{\citenamefont {Thouless}(1973)}]{thouless1973}%
      \BibitemOpen
      \bibfield  {author} {\bibinfo {author} {\bibfnamefont {D~J}\ \bibnamefont
      {Thouless}},\ }\bibfield  {title} {\enquote {\bibinfo {title} {Localization
      distance and mean free path in one-dimensional disordered systems},}\ }\href
      {\doibase 10.1088/0022-3719/6/3/002} {\bibfield  {journal} {\bibinfo
      {journal} {J. Phys. C: Solid State Phys.}\ }\textbf {\bibinfo {volume} {6}},\
      \bibinfo {pages} {L49--L51} (\bibinfo {year} {1973})}\BibitemShut {NoStop}%
    \end{thebibliography}
\end{document}